\documentclass{aa}  

\usepackage{soul}
\usepackage{amsmath}
\usepackage{booktabs}
\usepackage{multirow}
\usepackage{makecell}
\usepackage{enumitem}
\usepackage{array}
\usepackage{placeins}
\usepackage{dirtytalk}
\usepackage{xcolor}
\usepackage{comment}
\usepackage{lscape}
\usepackage[rightcaption]{sidecap}
\usepackage{threeparttable}
\usepackage{graphicx}
\usepackage{txfonts}
\usepackage{hyperref}
\sidecaptionvpos{figure}{c}
\begin{document}

   \title{Cocoon shock, X-ray cavities and extended Inverse Compton emission in Hercules~A: clues from {\it Chandra} observations}
   \titlerunning{Cocoon shock, X-ray cavities and extended Inverse Compton emission in Hercules~A}


   \author{F. Ubertosi
          \inst{1,2}\fnmsep\thanks{Contact: \texttt{francesco.ubertosi2@unibo.it}}
          \and
          Y. Gong\inst{3}
          \and 
          P. Nulsen\inst{4}
          \and
          J. P. Leahy\inst{3}
          \and
          M. Gitti\inst{1,5}
          \and
          B. R. McNamara\inst{6}
          \and
          M. Gaspari\inst{7}
          \and
          \\ M. Singha\inst{8,9,10}
          \and
          C. O'Dea\inst{11}
          \and
          S. Baum\inst{11}
          }

   \institute{Dipartimento di Fisica e Astronomia, Università di Bologna, via Gobetti 93/2, I-40129 Bologna, Italy,
         \and
             Istituto Nazionale di Astrofisica - Osservatorio di Astrofisica e Scienza dello Spazio (OAS), via Gobetti 101, I-40129 Bologna, Italy,
        \and
            Jodrell Bank Centre for Astrophysics, School of Physics \& Astronomy, University of Manchester, Oxford Rd., Manchester M13 9PL, UK,
        \and
            Center for Astrophysics $|$ Harvard \& Smithsonian, 60 Garden Street, Cambridge, MA 02138, USA,
        \and 
            Istituto Nazionale di Astrofisica - Istituto di Radioastronomia (IRA), via Gobetti 101, I-40129 Bologna, Italy,
        \and
            Department of Physics and Astronomy, Waterloo Centre for Astrophysics, University of Waterloo, Waterloo, ON N2L 3G1, Canada,
        \and
            Department of Physics, Informatics and Mathematics, University of Modena and Reggio Emilia, 41125 Modena, Italy,
        \and
            Astrophysics Science Division, NASA, Goddard Space Flight Center, Greenbelt, MD 20771, USA
        \and
            Department of Physics, The Catholic University of America, Washington, DC 20064, USA,
        \and
            Center for Research and Exploration in Space Science and Technology, NASA Goddard Space Flight Center, Greenbelt, MD 20771, USA,
        \and
           Department of Physics and Astronomy, University of Manitoba, Winnipeg, MB R3T 2N2, Canada.
            }

   \date{Received: September 30, 2024; Accepted: November 15, 2024}
%
 
  \abstract
  {}
   {We present a detailed analysis of jet activity in the radio galaxy 3C~348 at the center of the galaxy cluster Hercules~A. We aim to investigate the jet-driven shock fronts, the radio-faint X-ray cavities, the eastern jet, and the presence of extended Inverse Compton (IC) X-ray emission from the radio lobes.}
   {We use archival \textit{Chandra} observations to investigate surface brightness profiles extracted in several directions and to measure the spectral properties of the hot gas and of the non-thermal emission from the radio jet and lobes.} 
   {We detect two pairs of shock fronts: one in the north-south direction at 150 kpc from the center, and another in the east-west direction at 280 kpc. These shocks have Mach numbers of $\mathcal{M}=1.65\pm0.05$ and $\mathcal{M}=1.9\pm0.3$, respectively. Together, they form a complete cocoon surrounding the large radio lobes. Based on the distance of the shocks from the center, we estimate that the corresponding jet outburst is 90 -- 150 Myr old. We confirm the presence of two radio-faint cavities within the cocoon, misaligned from the main lobes, each approximately 100~kpc wide and 40–60 Myr old. A backflow from the radio lobes might explain why the cavities appear to be dynamically younger than the surrounding cocoon shock front. We also detect non-thermal X-ray emission from the eastern jet and from the large radio lobes. The X-ray emission from the jet is visible at 80~kpc from the AGN and can be accounted for by an IC model with a mild Doppler boosting ($\delta\sim2.7$). A synchrotron model could explain the observed radio-to-X-ray spectrum only for very high Lorentz factors $\gamma\geq10^{8}$ of the electrons in the jet. For the large radio lobes, we argue that the X-ray emission has an IC origin, with a 1~keV flux density of $21.7\pm 1.4 \,\text{(statistical)}\pm 1.3 \,\text{(systematic)}$~nJy. A thermal model is unlikely, as it would require unrealistically high temperature, density, and pressure for the gas in the lobes, along with a strong depolarization of the radio lobes, which are instead highly polarized. The IC detection, combined with the synchrotron flux density, suggests a magnetic field of $12\pm3$~$\mu$G in the lobes.}
   {}


   \maketitle
%

\section{Introduction} \label{sec:intro}
The power released by the active galactic nucleus (AGN) in the center of clusters of galaxies is the most promising explanation for the absence of run-away cooling \citep{1994ARA&A..32..277F} in the intracluster medium (ICM), and of extreme star formation in the  brightest cluster galaxy (BCG) \citep{2019SSRv..215....5W}. The power is provided by the accretion of cooled or cooling gas onto a supermassive black hole (SMBH) at the heart of the BCG \citep{1993MNRAS.263..323T,1997ApJ...484..602T}, creating a feedback loop that maintains the stability of the cool core \citep{2012ARA&A..50..455F,2007ARA&A..45..117M,2012NJPh...14e5023M,2012AdAst2012E...6G,2020NatAs...4...10G}.
\par In radio-mode or \emph{jetted} AGN \citep{2017NatAs...1E.194P}, the energy is mainly released in the form of jets of plasma \citep[for reviews, see][]{2019SSRv..215....5W,2020NewAR..8801539H,2022JApA...43...97S,2022JApA...43...85S}. The jets of radio galaxies can gradually dissipate as they slow down, terminating in diffuse radio lobes 
(\citealp[e.g.,][]{1974MNRAS.167P..31F} class I, hereafter FR I); alternatively, they can remain relativistic while propagating to the terminal shock at the hotspot, where the jets then inflate the high pressure cocoons which are viewed as radio lobes (\citealp[e.g.,][]{1974MNRAS.167P..31F} class II, hereafter FR II). While undergoing rapid expansion, the radio lobes of cluster-central AGN can push aside the gas creating X-ray cavities and driving shocks into the ICM  (e.g., \citep{2012ARA&A..50..455F,2007ARA&A..45..117M,2011MNRAS.411..349G,2019MNRAS.484.3376L,2022MNRAS.516.3750H,2022NatAs...6..109M,ubertosi23}. 
\par In a few cases, radio lobes of AGN in clusters can be associated with an excess, rather than a deficit, in the X-ray emission (e.g., \citealt{hardcastle2002,2002ApJ...580L.111I,2005ApJ...632..781I,2009ApJ...706..454I,2005ApJ...626..733C,2005MNRAS.363..649H,2009MNRAS.400..480K,2018MNRAS.478.4010D,2021ApJ...912...88G}). The emission mechanism has been found to be the inverse Compton (IC) scattering of photons -- primarily supplied by Cosmic Microwave Background (CMB) -- into X-ray photons on relativistic electrons with $\gamma\sim1000$ \citep{1979MNRAS.188...25H}. IC emission can serve as a probe of less energetic electrons and measure the electron density. In turn, combining the X-ray IC emission with radio emission intensities, it is possible to impose constraints on the magnetic field strength in the lobe (e.g., \citealt{feigelson1995,brunetti1999,hardcastle2002,2017MNRAS.470.2762M}).
\linebreak
\par In this article, we investigate the ICM around 3C~348, the fifth brightest extragalactic radio source at 178 MHz in the sky \citep{2013ApJ...771...38O}. 3C~348 resides in the Hercules~A galaxy cluster, located at a redshift $z=0.1549$ and with bolometric X-ray luminosity $L_X\sim5\times10^{44}$~erg/s \citep{1999A&A...350...25S,2004MNRAS.350..865G}. Possessing features from both FR I and FR II, namely two radio lobes extended over $\sim$400~kpc with a well-defined boundary but lacking hotspots, 3C~348 is generally categorized as an intermediate FR I/II source \citep[e.g.,][]{1991ApJ...379..141M,2002AJ....123.2312S,2003MNRAS.342..399G,2022A&A...658A...5T}. The eastern radio structure of Hercules~A is dominated by a jet gradually bending and diffusing into the lobe, while the western lobe features a distinct series of three rings \citep{1984Natur.308...43D,2003MNRAS.342..399G,2022A&A...658A...5T}. Recently, \citet{2022A&A...658A...5T} proposed that the rings were generated from a sequence of AGN outbursts each lasting over $0.4$~Myr. The difference between the two lobes is likely explained by considering the relativistic beaming of the jets combined with light-travel time delay across the inclined lobes, estimated to be $50^{\text{o}}$ to the line of sight (LOS), with the eastern lobe facing us \citep{2003MNRAS.342..399G,2004MNRAS.350..865G,2022A&A...658A...5T}. 
\par From the X-ray point of view, \citet{2010MNRAS.404.2018H} noted the presence of bright X-ray emission in the eastern lobe coincident with the radio jet, likely of synchrotron origin. \citet{2005ApJ...625L...9N} discovered two $\sim$80~kpc-wide X-ray cavities off the radio jets in positions nearly symmetric about the AGN, and a shock with Mach number $\simeq$1.65 at $58''$ from the AGN. \citet{2013ApJ...771...38O} proposed a scenario that cavities were generated by the previous epoch of activity $\sim$60~Myr ago releasing a mechanical energy of $3\times10^{61}$~erg \citep{2005ApJ...625L...9N}, before the AGN rotated about $\sim$65$^{\circ}$ to its current alignment about 20~Myr ago \citep{2013ApJ...771...38O}. 
\par This article focuses on the {\it Chandra} data of this galaxy cluster, to study the shocks and cavities, the X-ray and radio jets, and non-thermal emission associated with the jets and the lobes. 
We assume a flat $\Lambda$CDM cosmology with $H_0 =70\ \text{km}\ \text{s}^{-1}\ \text{Mpc}^{-1}$ and $\Omega_m=0.3$, giving a scale of $2.67\ \text{kpc}\ \text{arcsec}^{-1}$ for Hercules~A. Uncertainties are reported at 1$\sigma$ unless otherwise stated.

\begin{figure}[ht!]
    \centering
    \includegraphics[width=\linewidth]{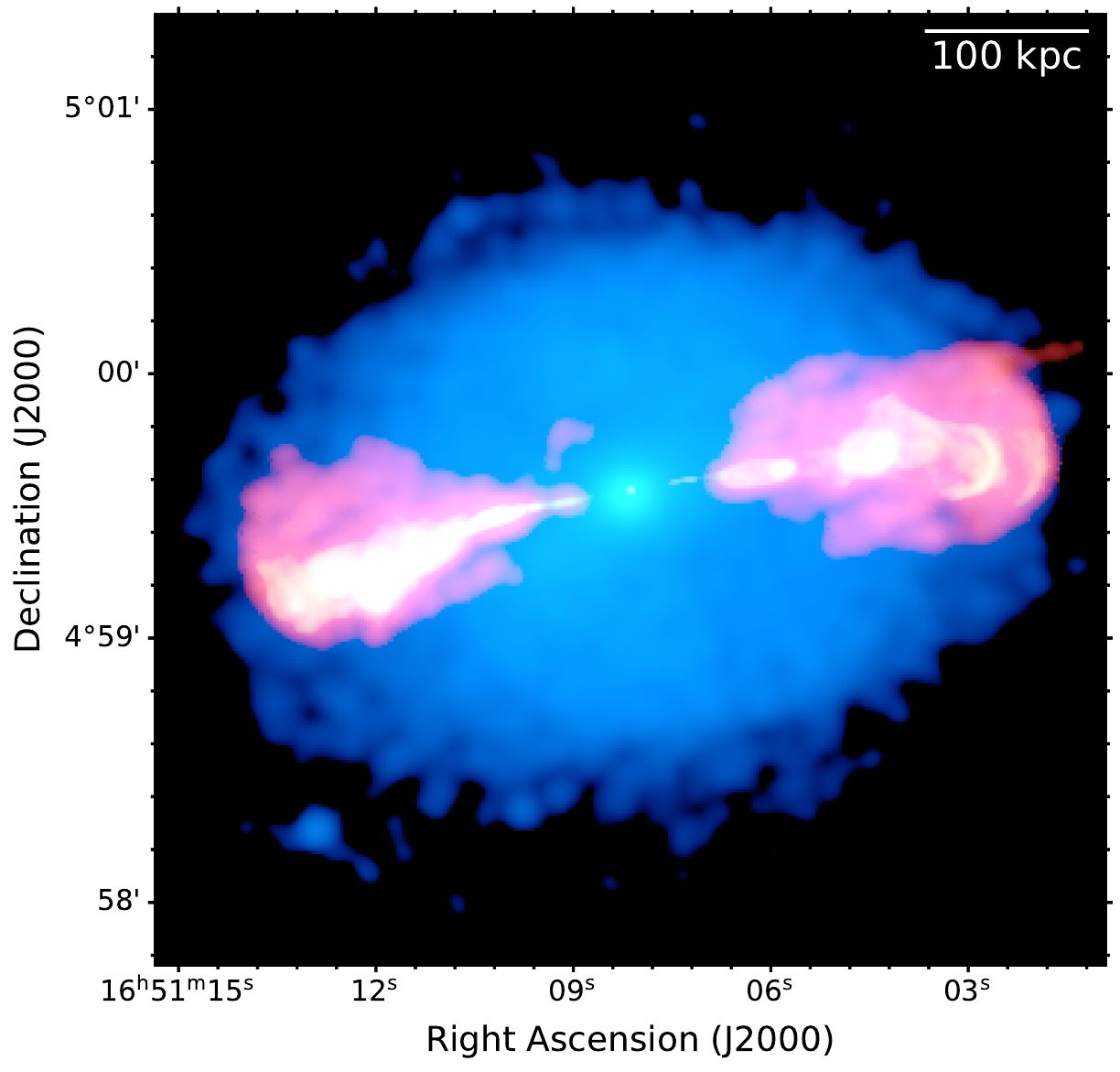}
    \caption{Large scale composite view of Hercules~A. Blue is \textit{Chandra} (X-ray) smoothed to a resolution of 5$''$, red and yellow are VLA (1.4 GHz, 4$''$ resolution, and 4.8 GHz, 1.5$''$ resolution, respectively). The field of view is a box of $4'\,\times$~4$'$, corresponding to about $640\,\text{kpc}\,\times$~640~kpc at the redshift $z = 0.1549$ of Hercules~A.}
    \label{fig:largeview}
\end{figure}
\section{Data reduction} \label{sec:style}
We reprocessed the three archival {\it Chandra} ObsIDs 1625 (15 ks), 5796 (50 ks), 6257 (50 ks) using CIAO-4.13 and CALDB-4.9.6. We used the longest observation (ObsID 5796) as reference to correct the astrometry. Background files were obtained from blank sky event files, normalized to the 9-12 keV count rate of the observations. The removal of background flares reduced the total exposure by $\sim$6.5\% to roughly 107 ks. We merged the event files using the \texttt{merge\_obs} script of \texttt{CIAO-4.14} to project events onto a common plane. From these, we produced a merged, exposure-corrected, background subtracted {\it Chandra} image in the 0.5 -- 7 keV band. In order to highlight substructures in the ICM, we smoothed the {\it Chandra} image with two Gaussian functions of 3$''$ (image \say{U3}) and 30$''$ (image \say{U30}) kernel size, respectively; then, we derived an unsharp masked image from the operation $(\text{U3}-\text{U30})/(\text{U3}+\text{U30})$. Notable features identified in the merged and unsharp images have then been investigated using \texttt{CIAO} and \texttt{Proffit}, while spectral fitting (in the 0.5 - 7 keV band) has been performed using \texttt{Xspec12.10}, selecting the table of solar abundances of \citet{asplund2009} and the C-statistic. Background spectra were extracted from blank-sky event files, and we have verified that there are no significant differences in our spectral results between using a local background spectrum (from an off-center region close to the edge of the field of view) and the blank-sky event files. An absorption model (\texttt{tbabs}) was always included to account for Galactic absorption, with the column density fixed at $N_{\text{H}} = 5.29\times10^{20}$ cm$^{-2}$ \citep{hi4pi2016}. The redshift was frozen to $z=0.1549$.
\par As complementary data, we consider the Very Large Array (VLA) observations at 1.4 GHz (A, B configuration), 4.9 GHz (B configuration), 8.5 GHz (C configuration) and 14.9 GHz (D configuration) archived at the NRAO VLA Archive Survey (NVAS\footnote{\url{http://www.vla.nrao.edu/astro/nvas/}.}). Such combination of frequencies and array configurations provides uniform sensitivity to structures on similar spatial scales, and matching angular resolution of about 1.5$''$. For imaging purposes, we also display radio contours from the 1.4 GHz VLA image (A+B+C configurations) presented in \citet{2003MNRAS.342..399G}, at a resolution of 1.4$''$.
\section{Results} \label{sec:results}

There are several notable features in the {\it Chandra} merged and unsharp masked images. As shown in Fig. \ref{fig:largeview} and \ref{fig:shockboundary}, a well-defined edge can be seen $\sim$1~arcmin north and south from the center, which had been identified by \citet{2005ApJ...625L...9N} as a shock front driven by the AGN activity. The ICM has an overall elliptical shape centered at the bright core, with its major axis rotated about $10^{\circ}$ counterclockwise from the east-west axis. The elliptical shape is aligned with the jet axis of the large radio lobes located to the east and west of the AGN, that reach a distance of about 1.5~arcmin ($\sim$240~kpc) from the center. Within the radio lobes there is significant X-ray emission, which further extends outwards, ending into two east and west edges at around $\sim$1.7~arcmin ($\sim$280~kpc) from the AGN. These peculiar features suggest that a whole cocoon shock surrounds the radio galaxy in Hercules~A, the boundary of which is traced by the north-south edges at 1~arcmin and the east-west edges at 1.7~arcmin. 
Additionally, regions of lower surface brightness are visible in the {\it Chandra} unsharp masked image of Fig. \ref{fig:shockboundary} in the northeast and southwest direction, corresponding to the X-ray cavities identified by \citet{2005ApJ...625L...9N}. Finally, we note the presence of a bright X-ray feature at $\sim$60$''$ east to the AGN, clearly associated with the radio jet and likely representing non-thermal emission. The eastern radio jet starts approximately aligned with the radio axis, and then deflects southward as it extends away from the AGN.

\subsection{Surface brightness analysis} \label{subsec:sbanalysis}

\subsubsection{The shock fronts}\label{subsubsec:sbprofiles}
\begin{figure*}[ht!]
    \centering
    \includegraphics[width=0.92\linewidth]{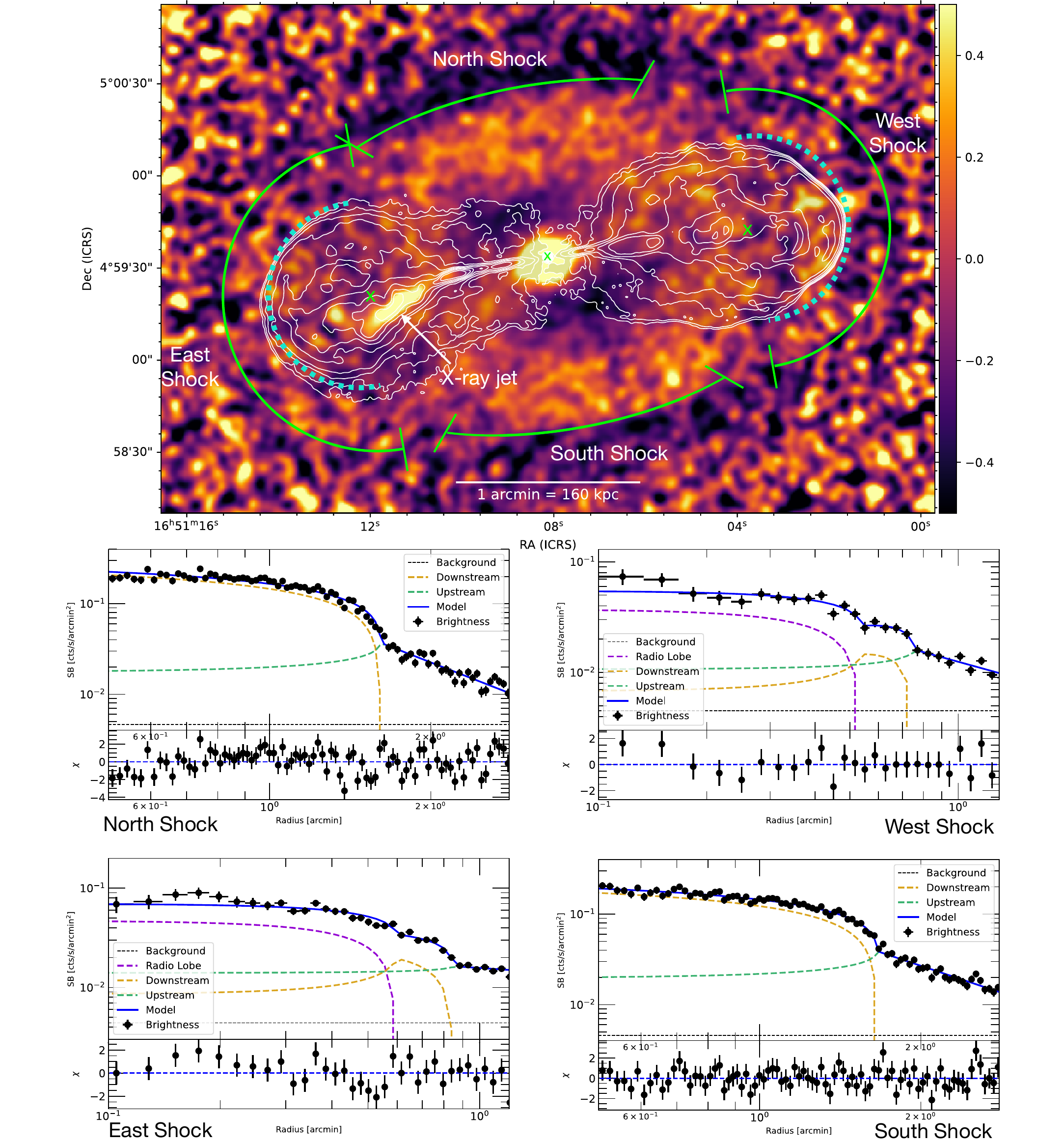}
    \caption{Analysis of surface brightness profiles across the shocks of Hercules~A. {\it Top panel:} {\it Chandra} unsharp masked image with 1.4~GHz ABC-configurations contours overlaid in white. Contours start at 3$\sigma_{\text{rms}} = 0.3$~mJy/beam and increase by a factor of 2. Green solid arcs show the position of the discontinuities and features identified in the surface brightness profiles centered in the green crosses. Cyan dotted arcs show the best-fit radius of the sphere of constant emission used to fit the surface brightness profile of the X-ray emission within the radio lobes. 
    {\it Bottom panels}: surface brightness profiles across the north, west, south, and east edges, fitted with a broken power-law model. The solid and dashed lines show the total model and the model components, respectively, projected along the line of sight. The black dotted line shows the background. For the north-south edges, the x-axis is relative to the major axis of the ellipse with ellipticity 1.75 and position angle 10$^{\circ}$. Best fit parameters are reported in Table \ref{tab:SBprofile-shocks}. For the east and west shocks, the profile was fit by combining the broken power-law model with a sphere of constant emission projected along the line of sight, to take into account the X-ray emission from the radio lobes. The best-fit profile shows the boundary of the radio lobes and of the shock front.
    }
    \label{fig:shockboundary}
\end{figure*}
To identify the exact position and magnitude of the surface brightness edges visible in the {\it Chandra} image, we analyzed surface brightness profiles extracted in specific wedges. The north-south edges at $\sim$1 arcmin from the center, first noted by \citet{2005ApJ...628..629N}, are the clearest feature in the X-ray image. 
We used two elliptical wedges with ellipticity 1.75 (we tested different values between 1.5 and 2.0, based on the morphology of the ICM, without finding significant differences) and position angle 10$^{\circ}$. The north and south wedges range between 60$^{\circ}$ and 150$^{\circ}$ (north edge), and between 230$^{\circ}$ and 320$^{\circ}$, respectively (angles are measured counterclockwise from west). Another set of edges is visible in the {\it Chandra} image, surrounding the large radio lobes of Hercules~A. For these east-west outer edges, we centered the surface brightness profiles in the geometrical center of the radio lobes. The circular wedges extend from $100^\circ$ to $280^\circ$ for the east outer edge, and from $280^{\circ}$ to $100^{\circ}$ for the west outer edge. The profiles are shown in Fig. \ref{fig:shockboundary}.
 \par We modeled the surface brightness profiles with a broken power-law model using \texttt{pyproffit} \citep{eckert2020}. This model has five parameters: the normalization $A_{s}$, the density jump $J$, the radial distance of the jump from the center $r_{J}$ (plotted along the major axis of the ellipse in Fig. \ref{fig:shockboundary} and reported along the minor axis in Table \ref{tab:SBprofile-shocks}), and the index of the inner and outer power-laws $\alpha_{1}$ and $\alpha_{2}$. 
The density jump is related to the Mach number $\mathcal{M}$ through the Rankine - Hugoniot conditions:
\begin{equation}
    \label{machn}
    \mathcal{M} = \left( \frac{3J}{4-J} \right)^{1/2}.
\end{equation}
The results of the fitting procedure are reported in Table \ref{tab:SBprofile-shocks}. The north and south edges are well described by density jumps of $\sim$1.9, at around $55''$ ($\sim$150~kpc) from the center. This is in good agreement with the results of \citet{2005ApJ...625L...9N}. The average Mach number of the north-south edges, determined from surface brightness modeling, is $\mathcal{M}_{\text{SB}} = 1.65\pm0.05$.
\begin{table*}
    \centering
    \footnotesize
    \caption{Surface brightness analysis of the shock fronts.}\label{tab:SBprofile-shocks}
    \renewcommand*{\arraystretch}{1.9}
    \begin{tabular}{l|c|c|c|c|c|c|c|c|c|c|c}
    \hline
      Edge &$\theta_{in}$&$\theta_{out}$& $J$ & $r_{J}$ & $\log{A_{s}}$ & $\alpha_{1}$ & $\alpha_{2}$ & $r_{l}$ &$\log{A_{l}}$ & $\chi^{2}/$ & $\mathcal{M}_{\text{SB}}$ \\
      &[$^\circ$]&[$^\circ$]&  & [arcmin]& [cts/s/ & & & & [cts/s/& d.o.f.&   \\
      &&&  & & arcmin$^{2}$] & & & & arcmin$^{3}$]& &   \\
        \hline
    North$^{a}$  & 60 & 150 & $1.95^{+0.05}_{-0.09}$ & $0.923^{+0.005}_{-0.003}$ & $-1.05^{+0.02}_{-0.02}$ & $0.26\pm0.06$ & $1.60\pm0.06$ & -- & -- &  1.451 (79)& $1.69^{+0.04}_{-0.08}$\\
    \hline
    South$^{a}$  & 230 & 320 & $1.85^{+0.08}_{-0.08}$ & $0.941^{+0.006}_{-0.006}$ & $-1.09^{+0.02}_{-0.02}$ & $0.24\pm0.04$ & $1.49\pm0.06$  &--  &--  & 0.915 (79) & $1.61^{+0.06}_{-0.06}$ \\
    \hline
    East  & 100 & 280 & $2.30^{+0.70}_{-0.43}$ & $0.870^{+0.061}_{-0.042}$ & $-1.45^{+0.06}_{-0.06}$ & $0.70\pm0.40$ & $0.62\pm0.16$  & $0.59\pm0.01$ & ${-1.40^{+0.02}_{-0.03}}$& 1.364 (28) & $2.06^{+0.70}_{-0.41}$\\
    \hline
    West  &280 & 100&  $2.02^{+0.51}_{-0.32}$ & $0.750^{+0.022}_{-0.036}$ & $-1.36^{+0.11}_{-0.15}$ & $-0.70\pm0.31$ & $1.05\pm0.10$ & $0.52\pm0.02$ & ${-1.46^{+0.04}_{-0.04}}$& 1.057 (18)& $1.85^{+0.40}_{-0.28}$ \\
    \hline
    \end{tabular}
\tablefoot{Best-fit results of the broken-power law fit to the surface brightness profiles shown in Fig. \ref{fig:shockboundary}. The north-south profiles are centered at the core of Hercules~A, while the east-west profiles are centered at the center of the corresponding radio lobe. (1) Name of the edge; (2 - 3) azimuthal limits of the sector used to extract the profile; (4) density jump; (5) distance of the discontinuity from the center of the profile; (6) normalization; (7) index of the inner (downstream) power-law; (8) index of the outer (upstream) power-law; (9) radius of the lobe-related emission; (10) emission per unit volume of the lobe related emission (Eq. \ref{eq:lobemodel}); (11) reduced $\chi^{2}$ (d.o.f); (12) Mach number, computed from the density jump using Eq. \ref{machn}.
\\$^{a}$ Reported distances are measured along the minor axis of the elliptical sectors with ellipticity 1.75 and position angle $10^{\circ}$ shown in Fig. \ref{fig:shockboundary} (note that surface brightness profiles are plotted along the major axis).}
\label{tab:shockfits}
\end{table*}
\par Special care has been taken when modeling the east and west outer profiles. The {\it Chandra} image reveals significant X-ray emission within the radio lobes, which may be due to IC emission. Indeed, if there was no X-ray emission from within the radio lobes, the surface brightness profile would turn downward there. Instead, as visible in Fig. \ref{fig:shockboundary}, the profiles turn upward, which requires the emission per unit volume to be greater from within the lobes than even in the shocked ICM. Therefore, we modified the above broken power-law model to include a contribution at smaller radii due to X-ray emission from within the lobe. {We assumed for geometrical construction that the lobe and the shock are concentric, and that the major axis of the cluster and the radio axis are aligned. While this is true in projection, other geometries may apply in 3D: the lobes are oriented at about 50$^{\circ}$ to the line of sight \citep{2003MNRAS.342..399G}, and the original ICM orientation, unmodified by shocks, may be different. However, given the difficulty in accurately measuring the original X-ray orientation due to the strong impact of the jets on the ICM, we prefer to assume for simplicity that the 2D alignment matches the 3D alignment, rather than introducing a subjective relative orientation between the jets and ICM and adding more free parameters to the model.} The lobe can be described by a sphere of constant emission per unit volume, $A_{l}$ (see also \citealt{2020ApJ...891..173S}) with radius $r_{l}$. This model is projected along the line of sight:
\begin{equation}
\label{eq:lobemodel}
    I_{l}(r) = A_{l}\times{2}\sqrt{r_{l}^{2} - r^{2}}, 
\end{equation}
The combined broken power-law $+$ lobe model has been fit to the surface brightness profiles in the east and west direction. The fitted profiles are shown in Fig. \ref{fig:shockboundary}, and the best-fit parameters are reported in Table \ref{tab:SBprofile-shocks}.
\par The best fit broken power-law indicates that the east-west edges are located at around $\sim$130~kpc from the center of the associated lobe, corresponding to $\sim$280~kpc from the center of Hercules~A (along the radio jet axis). Both edges have higher Mach number than the north-south ones, with an average $\mathcal{M}_{\text{SB}}=1.9\pm0.3$. The best-fit model for the sphere representing the lobe emission has a radius of $r_{l} = 94.5\pm1.6$~kpc for the east lobe and of $r_{l} = 83.3\pm3.2$~kpc for the west lobe. Each of the dashed cyan arcs on the X-ray image of Fig. \ref{fig:shockboundary} marks the radius $r_{l}$ for the respective component of the surface brightness model. It is noteworthy that the arcs each match the outer edge of the corresponding radio lobe. This is evidence that the X-ray emission and the radio emission from within the lobes are of related non-thermal origin. We further explore this possibility in Section \ref{subsec:lobesIC}.

\subsubsection{The X-ray cavities} 
\begin{figure*}[ht!]
\centering
\sidecaption
\includegraphics[width=0.7\linewidth]{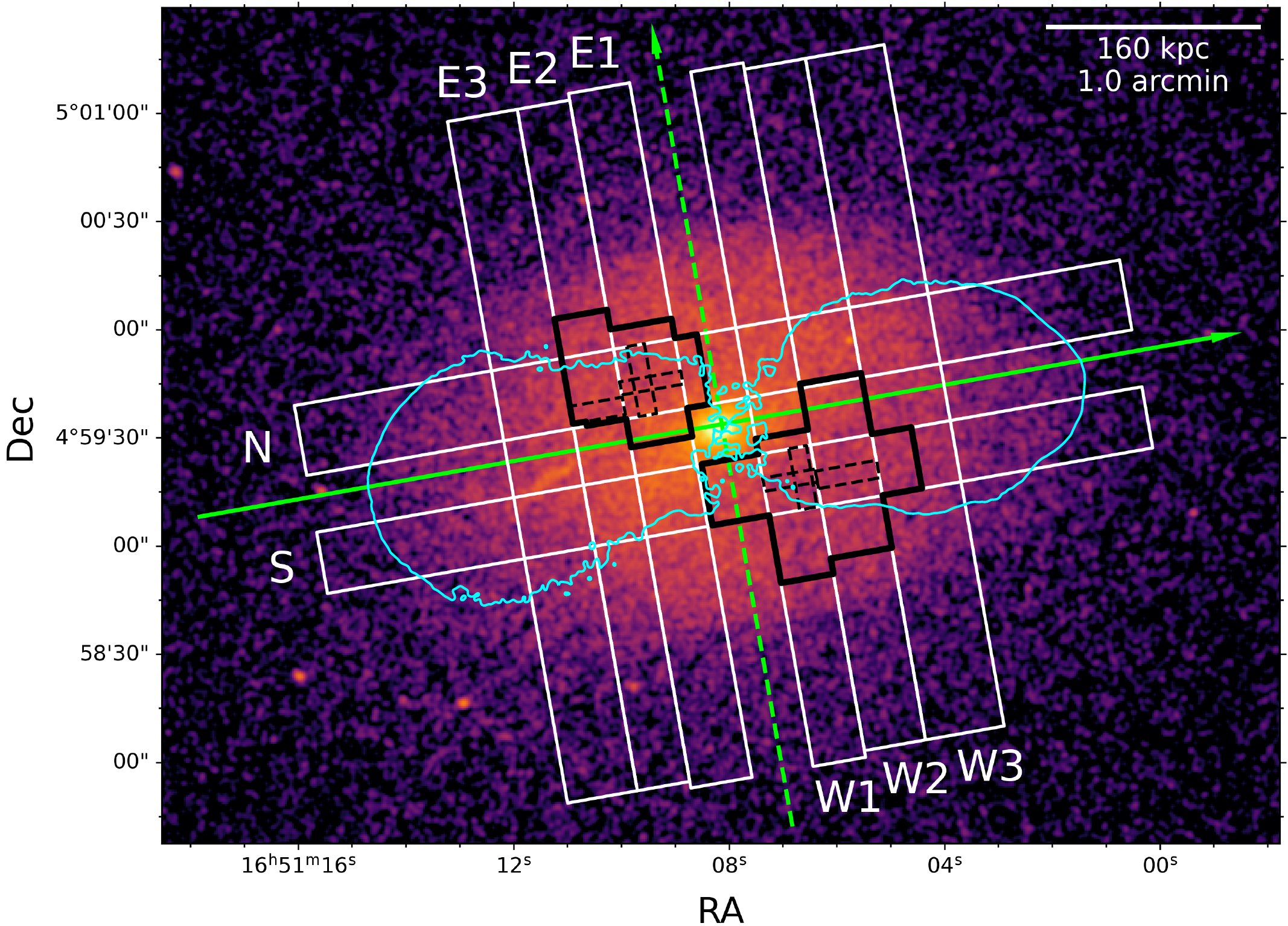}
\caption{Rectangular regions (white) used for the surface brightness profiles perpendicular and parallel to the radio axis, along with their labels. The solid black polygons encompass the cavity-related bins in each region. The dashed black rectangles mark the bin of the maximum deficit. The solid green line is aligned with the radio axis, where centers of the central bins of perpendicular cuts are positioned. The dashed green line, where the centers of the central bins of the parallel cuts is positioned, is perpendicular to the solid line. {The cyan contour from the 1.4~GHz ABC-configurations VLA is at 3$\sigma_{\text{rms}} = 0.3$~mJy/beam.}
\label{fig:cuts}}
\end{figure*}

\begin{figure*}
\centering
\includegraphics[width=0.9\textwidth]{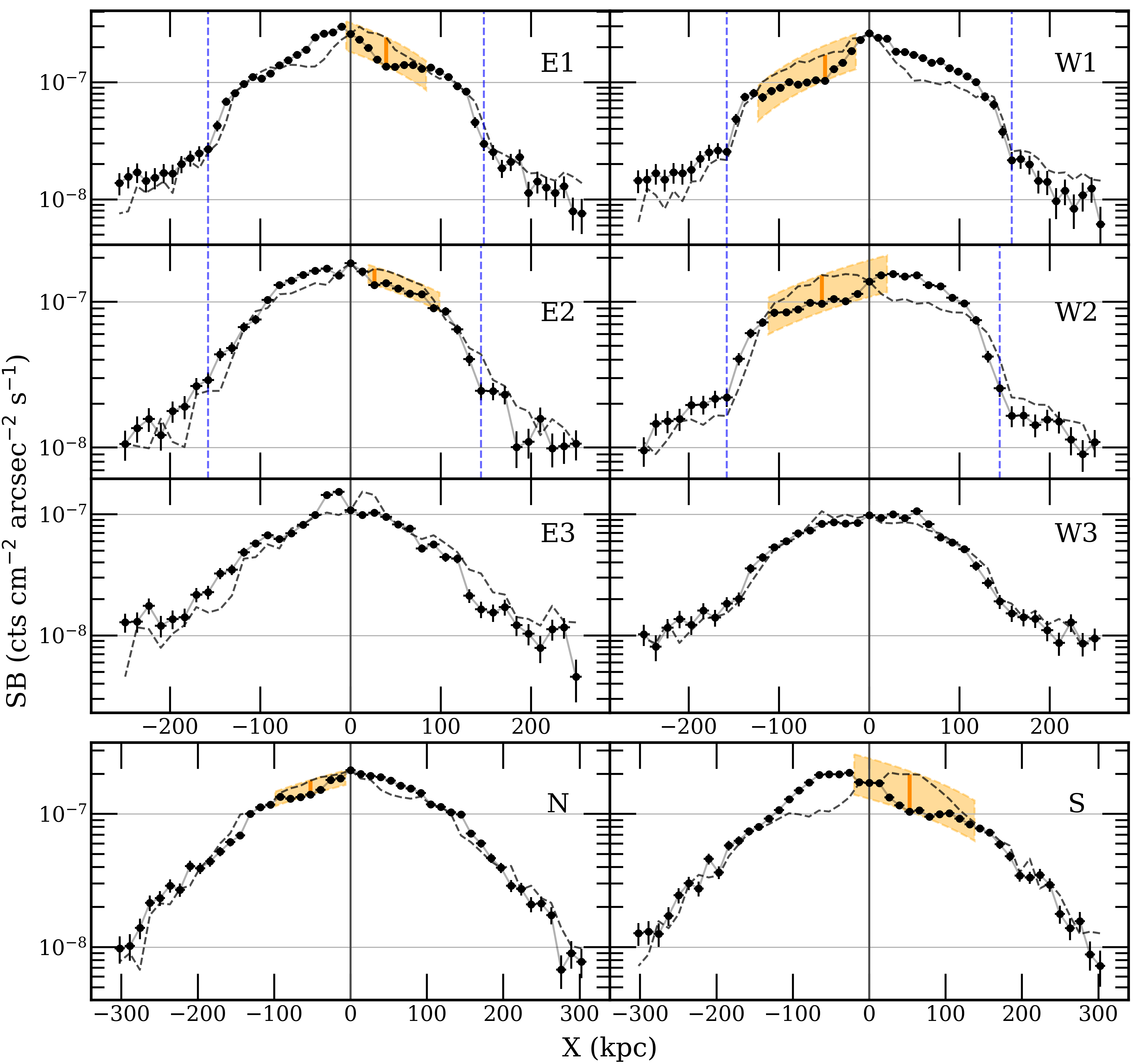}
\caption{Surface brightness cuts parallel and perpendicular to the radio axis. The labels correspond to those in Fig. \ref{fig:cuts}. The upper 3 rows are cuts from regions W1-W3 and E1-W3, where the positive x axis corresponds to the northern part of Hercules~A. The lowest row is cuts from regions N and S, where the positive x axis corresponds to the western part of Hercules~A. The dashed black lines are reflections of the solid lines about the central bin ($x=0$). The orange boxes illustrate the location of the cavities; within them the solid red lines mark the location of the greatest brightness deficit.}
\label{fig:perpcuts}
\end{figure*}

In order to determine the position and extent of the X-ray depressions, we analyzed surface brightness profiles extracted from rectangular regions that cut the ICM of Hercules~A in perpendicular and parallel directions to the radio axis (see Fig. \ref{fig:cuts} and Fig. \ref{fig:perpcuts}). To estimate the width of the cavities perpendicular to the radio axis, $w_{c}$, we considered the cuts E1, E2, E3, W1, W2, and W3. If no cavity was present, the surface brightness profiles would be symmetric about the radio axis; the cavity is, therefore, manifested as a deficit in surface brightness compared to the location reflected about the radio axis. If the cavity extends across the radio axis, it will cause the X-ray peak to lie on the opposite side to the cavity. Therefore, our criteria for cavity width $w_c$ estimation are: if the brightness peak is not on the same side as the X-ray depression or in the center, we consider one boundary of the cavity to be located at the surface brightness peak, otherwise it is located at the boundary of the first bin where the surface brightness at the mirrored location is larger; in both cases the outer edge of the cavity is located at the boundary of the first bin whose 1$\sigma$ error covers the surface brightness at the mirrored location. For each surface brightness cut, a deficit in X-ray surface brightness $\Delta S$ is estimated as the difference between the surface brightness in the bin with the largest depression, $S_0$, and the surface brightness at the location mirrored about the radio axis, $S_s$, i.e. $\Delta S=S_s-S_0$. We also denote with $x_c$ the distance of the bin with the largest depression from the radio axis (estimated as the location of the central bin). We report the results in Table \ref{tab:cavityproperty}. 
\par The presence of the cavities is confirmed by the asymmetry of the four profiles nearer the center (i.e. region E1, E2, W1, W2). We found no significant asymmetries in cuts E3 and W3 (the excess to the south of the center in E3 corresponds to the X-ray jet), indicating that the cavities do not extend to the E3 and W3 regions. We also find that the south-western cavity is larger than the north-eastern one, with an average $w_c \sim 120$~kpc for the south-western cavity and an average $w_c \sim 85$~kpc for the north-eastern one. 
\label{subsec:xraycavities}
\begin{table*}
\centering
\footnotesize
\renewcommand*{\arraystretch}{1.5}
\caption{Properties of the cavities from surface brightness cuts. }
\label{tab:cavityproperty}
\begin{tabular}{l|c|c|c|c|c|c|c|c|c|c}
\hline
{Cavity} & {Region} & {$x_c$} & {$l_c$} & {$w_c$} & {$w_g$} & {$S_0$} & {$S_s$} & {$\Delta S$} & {$\epsilon_{app}$} & {$d_c$} \\ 
 &  & [kpc] & [kpc] & [kpc] & [kpc] & [$10^{-7}$cts/s/ & [$10^{-7}$cts/s/ & [$10^{-7}$cts/s/ & [$10^{-10}$cts/s/ & [kpc]\\
 &  &  &  &  &  & cm$^{2}$/arcsec$^{2}]$ & cm$^{2}$/arcsec$^{2}]$ & cm$^{2}$/arcsec$^{2}]$ & cm$^{2}$/arcsec$^{2}/$ &  \\
  &  &  &  &  &  &  &  &  & kpc] &  \\
\hline
\multirow{3}{*}{SW} & W1 & $49\pm10$& -- & $108\pm14$ & $315\pm14$ & $1.03\pm0.07$ & $1.72\pm0.09$ & $0.69\pm0.12$ & $5.75\pm0.44$ & $120\pm22$ \\
& W2 & $53\pm13$ & -- & $131\pm19$ & $302\pm19$ & $0.97\pm0.06$ & $1.52\pm0.07$ & $0.55\pm0.09$ & $5.38\pm0.49$ & $103\pm19$ \\
& S & $53\pm13$ & $158\pm19$ & -- & -- & $1.04\pm0.06$ & $1.99\pm0.08$ & $0.95\pm0.10$ & -- & -- \\
\hline
\multirow{3}{*}{NE} & E1 &  $39\pm10$ & -- & $89\pm14$ & $305\pm14$ & $1.36\pm0.08$ & $2.42\pm0.11$ & $1.06\pm0.13$ & $8.20\pm0.56$ & $129\pm18$ \\
& E2 & $26\pm13$ & -- & $79\pm19$ & $302\pm19$ & $1.30\pm0.07$ & $1.69\pm0.08$ & $0.38\pm0.11$ & $5.67\pm0.46$ & $67\pm20$ \\
& N & $53\pm13$ & $92\pm19$ & -- & -- & $1.40\pm0.06$ & $1.78\pm0.07$ & $0.38\pm0.09$ & -- & -- \\
\hline
\end{tabular}
\tablefoot{(1) Location; (2) name of the surface brightness cut; (3) distance of the maximum deficit to the central bin (red vertical line in Fig. \ref{fig:perpcuts}); (4) length of the cavity along the radio axis; (5) width of the cavities perpendicular to the radio axis; (6) width of the shocked region perpendicular to the radio axis; (7-8) surface brightness value at the maximum deficit and at the location mirrored about the radio axis; (9) maximum surface brightness deficit; (10) average apparent emissivity at the location mirrored about the radio axis of the maximum deficit; (11) depth (i.e. width along the line of sight) of the cavity. The uncertainties for lengths are estimated as the bin width.}
\end{table*}
\par In order to estimate the lengths $l_c$ of the cavities along the radio axis, we considered the cuts parallel to the radio axis (region N and S). The lengths are estimated using the same criteria as the width. Again, we find significant asymmetries in these cuts, with the south-western cavity being larger than the north-eastern one. The locations of the cavity-related bins we selected are within the black polygons in Fig. \ref{fig:cuts}, along with the locations of the maximum deficit.
\par Overall, the north-eastern and south-western cavities are about $45 - 65$~kpc in radius. 
From the location of the bin with the largest deficit (the $x_c$ values reported in Table \ref{tab:cavityproperty}), we find that the north-eastern and south-western depressions are located at about 65~kpc and 70~kpc from the center, respectively (estimated as $\sqrt{x_{c}^2+l_c^2}$ using $x_c$ values in E1 and W1 because they host a larger depression compared with E2 and W2). Based on our estimates of $S_0$ and $S_c$, the cavities display maximum average deficits of 40\% (south-western one), and of 35\% (north-eastern one) with respect to the ambient medium. Therefore, the deep {\it Chandra} data confirm the presence of X-ray depressions in Hercules~A, which have the typical properties of X-ray cavities \citep{2007ARA&A..45..117M,2012NJPh...14e5023M}. Interestingly, they are located at about 45$^{\circ}$ from the axis of the radio jet. In Section \ref{sec:discussion} we further discuss the properties and role of the cavities in the context of AGN activity in Hercules~A.
\par Additionally, we try to evaluate the depth of the cavities along the line of sight from the deficits in surface brightness. For this purpose, we assume that the ICM of Hercules~A within the inner cocoon shock would be axially symmetric about the radio axis without the cavities. Judging from the symmetric shape of the gas (Fig. \ref{fig:largeview}) and from the profiles shown in Fig. \ref{fig:perpcuts}, this is a reasonable assumption. The apparent emissivity of the gas, namely the surface brightness per unit length (its integral along the column of gas projected onto each bin gives the surface brightness), is proportional to the intrinsic X-ray emissivity. We first estimate the apparent emissivity of the ICM at the location of the largest deficit in a cavity mirrored about the radio axis, $\epsilon_{app}\approx S_s/d_s$. Here, $d_s$ is the depth along the line of sight of the ICM cocoon, which is estimated as $d_s=\sqrt{w_g^2-4x_c^2}$. The diameter of the cocoon shock, $w_g$, is estimated as the distance between the northern and southern edges, which can be measured from the profile of Fig. \ref{fig:perpcuts} as the distance between the bins where the surface brightness starts to increase steeply (the point where the rectangles cut the north and south edges). Assuming that there is no X-ray emission from within the cavity, using the maximum deficit in X-ray surface brightness we can estimate the depth of the cavity along the line of sight: $d_c=d_s(\Delta S/S_s)=\Delta S/\epsilon_{app}$, on the assumption that the emission per unit volume within the shocked ICM is uniform and that the background emission from inside and outside the cavity is negligible -- not a terrible approximation from the surface brightness profiles. Inevitably, this estimate is approximate. The X-ray emission from outside the edges is $\sim$10\% the central brightness judging from the surface brightness profiles, and alone results in a $\sim$10\% underestimations of $d_c$. 
{There might also be IC emission from the cavities. However, 74 MHz and 144 MHz radio observations  \citep{2005MNRAS.358.1061G,2022A&A...658A...5T} showed that the amorphous radio emission brightness in the region of the X-ray cavities is at least 5 times less than that from the lobes. Therefore, any X-ray IC brightness $S_\text{cav}$ from the cavities would also be at least five times lower than the IC brightness at the radio lobes (assuming similar magnetic field strengths), as the X-ray and radio emissions arise from the same population of electrons with $\gamma\sim10^3$. Since the lobe IC X-ray brightness $A_l\cdot r_l$ is only about 20\% of the shocked ICM brightness as reported in Table \ref{tab:shockfits}, then the putative X-ray IC brightness from the cavities would be $S_\text{cav}\lesssim S_0/25$. Thus, neglecting the possible $S_\text{cav}$ produces systematic underestimations of $\lesssim10-20$\% depending on $\Delta S/S_s$.}
\par The results are listed in Table \ref{tab:cavityproperty}. For both the south-western cavity and the north-eastern cavity, $d_c>w_c$ in E1 and W1, and $d_c<w_c$ in E2 and W2. The 1$\sigma$ confidence ranges of $d_c$ and $w_c$ overlap except in E1. The difference may be dominated by the systematic uncertainties. In additions to the underestimations mentioned above, the gas the cavities excavated, closer to the center of Hercules~A, is brighter than that of the peripheral regions, leading to a larger surface brightness depression than what it is for a uniformly-emitting gas in our assumptions, and hereby an overestimation of $d_c$. We express fiducial values of the sizes of the cavities as the average of different cuts, which are $\sim$90~$\times~80\times100$~kpc (length by width by depth) and $\sim$160~$\times~120\times110$~kpc for the north-eastern and south-western cavities, respectively. Given the crudeness of our analysis, we argue that the width and depth of both cavities are consistent with each other. The south-eastern cavity is likely larger than the north-western cavity.

\subsection{Spectral analysis}
\subsubsection{Thermodynamic properties of the shock fronts}\label{subsec:specshocks}
To measure the spectral properties of the detected surface brightness edges, we extracted the spectra of three concentric regions: the first region is a wedge containing the downstream side of the discontinuity ($r\leq r_{J}$); the second is a wedge containing the upstream side of the discontinuity ($r>r_{J}$); a third outer wedge extending to the edge of the {\it Chandra} field of view is used to deproject the spectra. The bin widths of the first two regions were chosen to avoid the inclusion of thermal emission far from the jump: while selecting larger regions would increase the number of counts, it may also lead to smearing thermodynamic gradients. The regions chosen for the analysis of the north-south and east-west discontinuities are shown in Fig. \ref{fig:specanalysisregions}. Spectra were fitted with a \texttt{projct$\ast$tbabs$\ast$apec} model using the Cash statistic in the 0.5 -- 7 keV band to measure the deprojected temperature and density, which were combined to derive the pressure jump across each edge. The details of the spectral analysis are reported in Table \ref{tab:specfronts}.
Below we present the results for the discontinuities in Hercules~A.
\paragraph{North-south shock fronts}\mbox{}
We combined the regions defined by the northern and southern wedges shown in Fig.~\ref{fig:specanalysisregions}. Since the edges identified from the surface brightness modeling are concentric (see Table \ref{tab:SBprofile-shocks} and Section \ref{subsubsec:sbprofiles}), and their symmetry about the center strongly supports a common origin, we jointly fitted the spectra extracted from the two sets of wedges. As reported in Table \ref{tab:specfronts}, we measured a downstream temperature of $kT_{1}=5.68_{-0.33}^{+0.29}$~keV and an upstream temperature of $kT_{2}=4.01_{-0.21}^{+0.19}$~keV. This corresponds to a temperature jump of $kT_{1}/kT_{2} = 1.42_{-0.11}^{+0.10}$. The density jump is $n_{e,1}/n_{e,2} = 2.00_{-0.07}^{+0.07}$, which is in agreement with that found from the surface brightness modeling of $J = 1.90\pm0.06$ (average of the north and south jumps), and the pressure jump is $p_{1}/p_{2} = 2.85_{-0.24}^{+0.23}$. 

\paragraph{East-West shock fronts}\mbox{}
The east-west edges are not centered on the AGN, and the radius of the density jump, $r_{J}$, differs between the east and west surface brightness profiles (see Table \ref{tab:SBprofile-shocks}). The lobe radius, which sets the inner limit for the spectral extraction, also differs between the east and west. These differences are not dramatic, but they prevent us from jointly fitting a deprojected thermal model to the spectra of the east and west discontinuities. Therefore, we proceed with separate deprojected models of the two edges. The results are as follows:
\begin{itemize}[leftmargin=*]
    \item For the eastern edge, as reported in Table \ref{tab:specfronts}, we measured a downstream temperature of $kT_{1}=7.74_{-1.15}^{+1.63}$~keV and an upstream temperature of $kT_{2}=5.72_{-0.94}^{+1.33}$~keV. This corresponds to a temperature jump of $kT_{1}/kT_{2} = 1.35_{-0.39}^{+0.63}$. The density jump is $n_{e,1}/n_{e,2} = 1.74_{-0.14}^{+0.15}$,  
    and the pressure jump is $p_{1}/p_{2} = 2.36_{-0.69}^{+1.14}$. 
    \item For the western edge, we measured a downstream temperature of $kT_{1}=6.84_{-1.73}^{+2.70}$~keV and an upstream temperature of $kT_{2}=4.14_{-0.87}^{+1.06}$~keV. This corresponds to a temperature jump of $kT_{1}/kT_{2} = 1.65_{-0.67}^{+1.27}$. The density jump is $n_{e,1}/n_{e,2} = 2.09_{-0.22}^{+0.24}$, 
    and the pressure jump is $p_{1}/p_{2} = 3.46_{-1.43}^{+2.77}$. 
\end{itemize}
Overall, both the east and west surface brightness edges show the typical properties of shock fronts. Their Mach numbers are slightly higher than the typical Mach $\mathcal{M}\leq1.5$ of shock fronts in other cool core clusters \citep{2019MNRAS.484.3376L,ubertosi23}, but are still in the regime of weak shocks within uncertainties (for comparison, the shock around the southern lobe of Centaurus~A has $\mathcal{M}\sim8$, \citealt{2009MNRAS.395.1999C}).
Clearly, deeper X-ray observations would be needed to robustly confirm the associated temperature jumps, which we could only loosely constrain given the large uncertainties. 
\par We also note that the temperature jumps of the four edges determined from the spectral analysis, in the range $1.42\lessapprox T_{J}\lessapprox1.65$, are smaller than those expected from the Mach number $\mathcal{M}_{\text{SB}}$. Given the relation:
\begin{equation}
    T_{J} = \frac{5\mathcal{M}^{4}+14\mathcal{M}^{2}-3}{16\mathcal{M}^{2}},
\end{equation}
and the Mach numbers reported in Table  \ref{tab:SBprofile-shocks}, we would expect temperature jumps in the range $1.6\lessapprox T_{J}\lessapprox2.2$. This mismatch is likely caused by the lower spatial resolution of the temperature measurements with respect to the surface brightness profile, which smooths the thermodynamic gradient and makes the outcome more sensitive to preshock conditions.

\begin{figure*}[ht!]
    \centering
    \includegraphics[width=\linewidth]{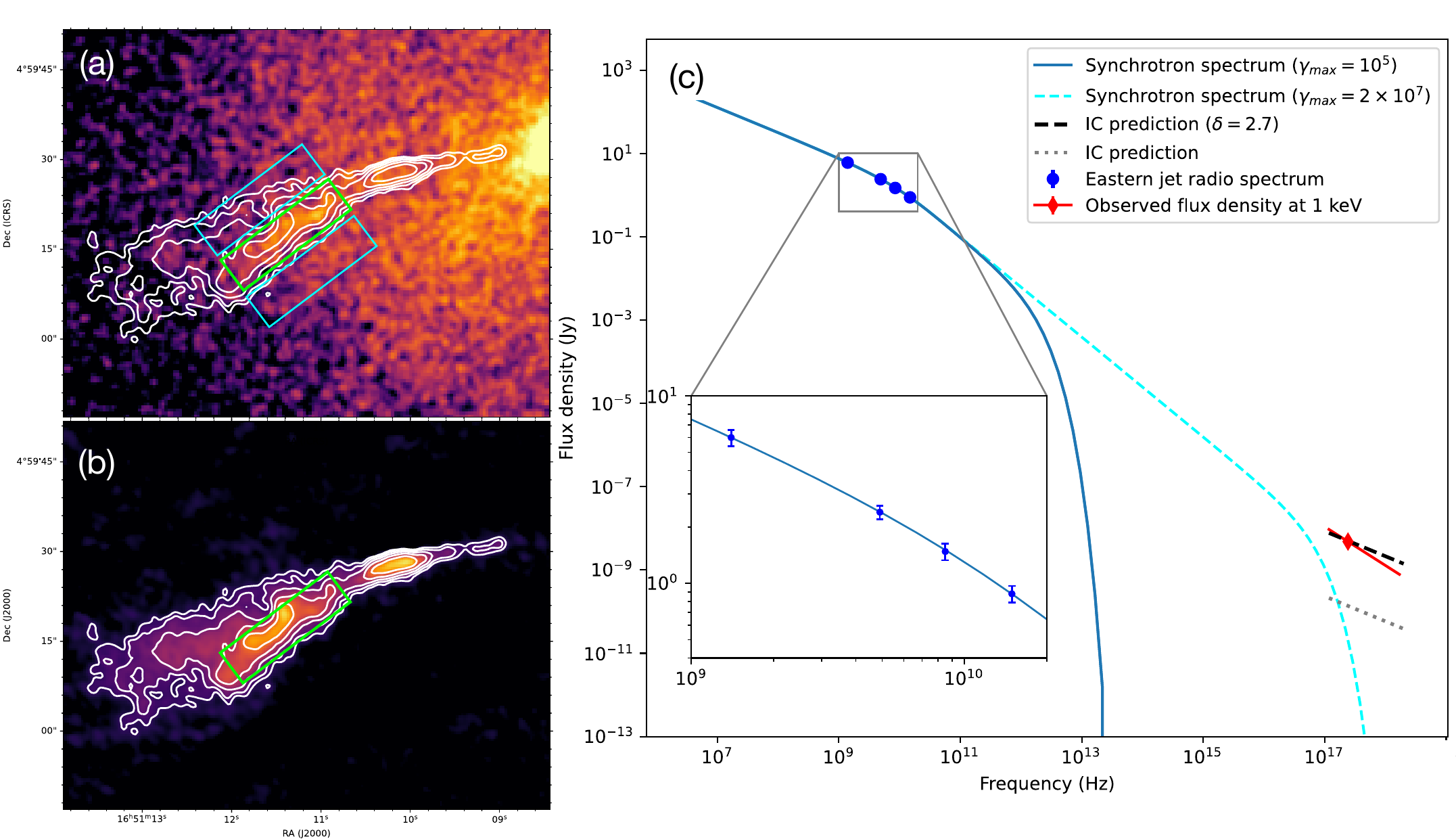}
    \caption{The eastern X-ray jet. \textit{Panel (a):} 0.5 -- 7 keV {\it Chandra} image with 14.9~GHz radio contours overlaid in white. The contours start at 3$\sigma_{\text{rms}} = 0.6$~mJy/beam and increase by a factor of 2. The green box is the region of enhanced X-ray emission along the eastern radio jet. Cyan boxes show the background extraction regions. The spectral analysis of its emission is reported in Section \ref{subsubsec:xrayknot}. \textit{Panel (b):} 14.9~GHz VLA image of Hercules~A, with contours defined as in the left panel. The beam FWHM is $1.8''\times 1.4''$. \textit{Panel (c):} radio to X-ray spectrum of the eastern X-ray jet within the green box. The blue points are the flux densities at 1.5, 4.9, 8.5 and 14.9 GHz, while the red diamond is the measured flux density at 1~keV. The blue solid and cyan dashed lines represent the synchrotron model describing the radio spectrum for $\gamma_{max} = 10^{5}$ and $\gamma_{max} = 2\times10^{7}$, respectively. The black dashed and grey dotted lines shows the prediction for IC of the CMB from the electrons powering the synchrotron emission, with and without Doppler boosting   
    (see Sect. \ref{subsubsec:xrayknot} and \ref{subsubsec:eastjetmodels} for details).}
    \label{fig:easternjet}
\end{figure*}
\subsubsection{The eastern X-ray jet} \label{subsubsec:xrayknot}
The thin X-ray feature in the eastern lobe first appears at $\sim$80~kpc from the center and is strongly correlated with the radio jet (Fig.~\ref{fig:shockboundary} and \ref{fig:easternjet}). The available {\it Chandra} observations collect $\sim$850 net counts in the 0.5 -- 7 keV band (using a local background to subtract the cluster emission) over a rectangle of length 20$''$ and width 6$''$ covering the eastern X-ray jet (shown in Fig. \ref{fig:easternjet}a). Fitting its spectrum with an absorbed power-law model, we measure a photon index of $1.36\pm0.25$ (3$\sigma$ error). The C-statistic is 348.46/425. The fit improves by adding an intrinsic absorption component, with a best-fit column density of $4.9^{+0.3}_{-0.2}\times10^{21}$ cm$^{-2}$. In this case, the power-law index is $1.9\pm0.5$ (3$\sigma$ error), and the C-statistic is 343.48/424. The addition of the intrinsic absorption yields an F statistic value of 6.15 and probability 1\%, indicating that this component is statistically required. \\We also tested whether a thermal \texttt{apec} model may provide a better fit to the X-ray emission from the jet. This may be the case if thermal plasma has been entrained by the radio jet. From this test we measure a temperature of $kT = 14.9^{+44}_{-8.9}$ keV (3$\sigma$ error). The abundance is not constrained, so it is fixed at 0.3 solar. The C-statistic is 347.31/425. Again, the fit improves when adding an intrinsic absorption component with a column density of $3\times10^{21}$ cm$^{-2}$. In this case, the temperature is $kT = 8.3^{+18}_{-2.5}$ keV (3$\sigma$ error), and the C-statistic is 343.35/423. 
Overall, the intrinsically absorbed power-law may represent the best-fit in terms of reduced C-stat over the thermal model. This is also suggested by the poorly unconstrained parameters of the \texttt{apec} component: a temperature of $\sim$10~keV seems unlikely, given that the ambient gas has a temperature of $\sim$5~keV; furthermore, the normalization ($2.9\times10^{-5}$) would imply unreasonably high gas density $n_e\sim0.03$~cm$^{-3}$ and pressure $p\sim8\times10^{-10}$~erg~cm$^{-3}$. These are an order of magnitude higher than the ambient density and pressure, and even higher than the central density and pressure of Hercules~A.

\subsubsection{IC emission within the radio lobes}\label{subsec:lobesIC}
The lack of evident X-ray cavities associated with the radio lobes of Hercules~A has been noted before (\citealt{2010MNRAS.404.2018H,2013ApJ...771...38O}). The X-ray cavities may be masked by IC X-ray emission from the radio lobes, as in Cygnus~A \citep{2018ApJ...855...71S,2020ApJ...891..173S}. Our analysis of Section \ref{subsubsec:sbprofiles} corroborates the idea that non-thermal emission is causing the lobes to shine in X-rays and to mask any underlying cavities.
\par Previously, \citet{2010MNRAS.404.2018H} investigated the presence of X-ray emission from within the radio lobes of Hercules~A. Using {\it Chandra} data (OBSIDs 5796 and 6257 only), they found a 99\% upper limit on IC emission of 38~nJy for the two radio lobes combined. Here, we leverage all the available {\it Chandra} exposures and the information retrieved from the surface brightness modeling to address again this point. For completeness, we note that ICM thermal emission unlikely explains the presence of X-ray emission inside the lobes. As detailed in Appendix \ref{app:thermlobes}, fitting a thermal model to the spectra of the lobes we find unlikely high best-fit temperatures and densities. Even more, we demonstrate that if the lobes were filled with thermal plasma they should be completely depolarized, which is contrary to observational evidences provided by \citet{2003MNRAS.342..399G}.
\par We extracted the spectrum of the eastern and western radio lobes from the green half-circles shown in Fig. \ref{fig:specanalysisregions}, that have radii of $r_{l} = 35.4'' = 94.5$~kpc and $r_{l} = 31.2'' = 83.3$~kpc for the eastern and western lobes, respectively. We excluded the region containing the X-ray jet, which would bias our results towards a detection of non-thermal emission. As noted by \citet{2010MNRAS.404.2018H}, the background must be carefully chosen when searching for faint non-thermal emission embedded in a relatively X-ray brighter halo. The X-ray emission from the lobes of Hercules~A is a combination of the gas projected along the line of sight on top of the radio lobes, and the X-ray emission within them. We considered different approaches to fit the spectra: 
\begin{enumerate}[leftmargin=*]
    \item We adopted as background the shell of shocked gas immediately outside the radio lobe, whose spectrum accounts for the whole emission projected onto the lobe. We obtain 1565 net counts after background subtraction for the eastern lobe, and 971 net counts after background subtraction for the western lobe. Fitting the spectra with an absorbed power-law model, we find, for the outer portion of the eastern lobe, a photon index $\Gamma = 1.54\pm0.07$ and a flux density at 1 keV of $14.3\pm0.6$~nJy (C-stat/D.o.f. = 256.8/267). For the outer portion of the western lobe we find a photon index $\Gamma = 1.58\pm0.08$ and a flux density at 1 keV of $8.9\pm0.5$~nJy (C-stat/D.o.f. = 199.8/218). Thus, the combined non-thermal emission from the radio lobes equals $23.2\pm1.1$~nJy.
    \item As an alternative approach, we subtracted the blank-sky background from the source spectrum. Since the blank-sky does not account for thermal emission within or projected onto the lobes,  
    we fitted the lobe spectrum with a \texttt{tbabs$\ast$(apec$+$po)} model. The temperature and abundance of the thermal model were frozen to $kT = 4.1$~keV and $Z=0.5$~Z$_{\odot}$, obtained from an off-lobe region extending from $r_{l}$ to the edge of the field of view (similar to \citealt{2010MNRAS.404.2018H}; this abundance is also typical of cool cores, see e.g., \citealt{2009A&A...508..565D}). The normalization of the \texttt{apec} and \texttt{po} components and the photon index of the \texttt{po} component were left free to vary. For the eastern lobe we find a photon index $\Gamma = 1.56\pm0.09$ and a flux density at 1 keV of $14.9\pm1.4$~nJy (C-stat/D.o.f. = 217.9/227). 
    For the western lobe we find a photon index $\Gamma = 1.49\pm0.12$ and a flux density at 1 keV of $5.7\pm1.8$~nJy (C-stat/D.o.f. = 180.1/185). 
    The combined non-thermal emission from the radio lobes equals $20.6\pm1.9$~nJy.
\end{enumerate}
These results are consistent with each other, and are in agreement with the upper limit of 38~nJy found by \citet{2010MNRAS.404.2018H}. We note that \citet{2010MNRAS.404.2018H} used a larger source region (approximately twice as large), which also contains the radio emission within the inner 1 arcmin of Hercules~A. Thus, our result may underestimate the total IC emission. However, the majority of non-thermal emission is likely related to the brightest part of the radio lobes. Further, any contribution from within the innermost arcmin would be hard to constrain, given that the emission in those areas is dominated by the thermal cluster emission, and that the IC flux is likely lower. 
\par The normalization of the X-ray surface brightness across the radio lobes that we obtained in Section \ref{subsubsec:sbprofiles} can be used as a third method to estimate the flux density of the lobe X-ray emission. The count rate of each lobe is given by $A_{l}\times V_{l}$, where $V_{l}$ is the volume of the half-lobe region. For the eastern lobe, $V_{l} = 0.42$~arcmin$^{3}$ and the count rate is $(1.47\pm0.09)\times10^{-2}$ cts~s$^{-1}$. For the western lobe, $V_{l} = 0.29$~arcmin$^{3}$ and the count rate is $(1.18\pm0.06)\times10^{-2}$ cts~s$^{-1}$.
Assuming a power-law model with photon index 1.5 and taking into account the {\it Chandra} response\footnote{In PIMMS, \url{https://cxc.harvard.edu/toolkit/pimms.jsp}.}, these translate into flux densities at 1~keV of $11.8\pm0.7$~nJy and $9.5\pm0.5$~nJy for the eastern and western lobes, respectively. The combined flux density from the radio lobes equals, in this case, $21.3\pm1.2$~nJy. Since these are of the order of the spectrally-determined fluxes, we conclude that the surface brightness and spectral analyses provide comparable results. 
\\The comparison between the three methods can provide an estimate of systematic uncertainties. We express our fiducial value for the 1 keV flux density as the average between the methods, with the associated statistical uncertainty and the systematic uncertainty given by the dispersion, that is $21.7 \pm 1.4 \pm 1.3$~nJy.  



\section{Discussion}\label{sec:discussion}
\subsection{History of AGN activity in Hercules~A}\label{subsec:histAGN}
The unclear connection between the large radio lobes and the pair of radio-faint X-ray cavities in Hercules~A is puzzling. Previously, \citet{2013ApJ...771...38O} speculated that the cavities represent an older outburst with respect to the present large-scale jets. Specifically, \citet{2013ApJ...771...38O} estimated a dynamical age for the radio galaxy in Hercules~A of roughly 20~Myr, assuming a lobe expansion speed of 0.03$c$. However, for a plasma temperature between 4 and 6 keV (compatible with the ICM of Hercules~A), such an expansion speed would have generated a $\mathcal{M}\sim7-9$ shock, that would be clearly visible in the {\it Chandra} images. Thus, it is likely that the true age of 3C~348 is larger than 20~Myr. 
\par The shock fronts in the ICM can be used to infer the timescales of AGN activity in Hercules~A. Specifically, following e.g., \citet{randall11,ubertosi23}, we assume for simplicity that the shock front has propagated from the center of the BCG at a speed $v_{sh} = \mathcal{M}\times c_{\text{s}}$, where $c_{\text{s}}=\sqrt{\gamma kT/(\mu m_{\text{p}})}$ 
is the upstream sound speed. The shock age is thus given by:
\begin{equation}
    \label{eq:shockage}
    t_{\text{sh}} = \frac{r_{\text{J}}}{\mathcal{M}\,c_{\text{s}}}.
\end{equation}
For the E-W shock fronts, located at 287 and 294 kpc from the AGN and with Mach numbers given in Table~\ref{tab:SBprofile-shocks}, we consider the upstream temperature reported in Table~\ref{tab:specfronts} to derive $c_{\text{s}}$. We find $v^{E}_{sh}=(2.4\pm0.9)\times10^{3}$~km/s and $v^{W}_{sh}=(1.9\pm0.5)\times10^{3}$~km/s. The E-W shock fronts are located at the tip of the radio lobes, thus the above velocities may be interpreted as the average expansion speed of the lobes. From Equation \ref{eq:shockage} we obtain $t^{E}_{\text{sh}}=116\pm43$~Myr and $t^{W}_{\text{sh}}=152\pm38$~Myr. 
For the N-S shock front at 150 kpc from the center and with Mach numbers given in Table~\ref{tab:SBprofile-shocks}, we consider the upstream temperature reported in Table~\ref{tab:specfronts} to measure $v^{NS}_{sh}=(1.7\pm0.1)\times10^{3}$~km/s and $t^{NS}_{\text{sh}}=88\pm4$~Myr. As the E-W and N-S shock fronts are part of the single elliptical cocoon front that surrounds the radio galaxy, the above estimates provide an approximate range for the large-scale outburst age of $90\lessapprox t_{sh} \lessapprox 150$~Myr, indicating the magnitude of systematic uncertainties. For comparison, recent simulations by \citet{perucho2023} yield a cocoon shock size of $\sim$280~kpc after about 152~Myr, which is consistent with our results. 
It is important to note that the method used here may overestimate the shock age. Initially, the shock likely had a higher Mach number, which decreased rapidly over time (in a typical Sedov-Taylor point explosion, the Mach number decays as $\mathcal{M}\propto t^{-3/5}$). 
As a result, using Equation \ref{eq:shockage} to estimate the shock age may lead to a modest overestimation, by around 20\%, as reported by \citet{randall11,ubertosi23}. However, if the cocoon shock had a non-negligible inclination along the line of sight, projection effects could cause an underestimation of the shock radius, and thus the shock age given by Equation \ref{eq:shockage}. The combination of these opposite effects is likely captured within the shock age range of 90–150 Myr mentioned above.
\par For completeness, we can estimate the jet energetic and power associated with the large cocoon shock front. To compute the shock energy we consider that (e.g. \citealt{2001ApJ...557..546D,randall11}):
\begin{equation}
    \label{eq:shockenergymach}
    E_{\text{sh}} = \frac{3}{2} \,p_{\text{up}} \times V \times \left(\frac{5\mathcal{M}^{2}-5}{4}\right),
\end{equation}
where $p_{\text{up}}$ is the pre-shock (upstream) pressure, $V$ is the shocked volume, and $\mathcal{M}$ is the Mach number. Using the average upstream pressure $p_{\text{up}} = 2.1\pm0.5\times10^{-11}$~erg~cm$^{-3}$ (Table \ref{tab:specfronts}) and average Mach number $\mathcal{M} = 1.78\pm0.18$ (Table \ref{tab:SBprofile-shocks}), and assuming a prolate ellipsoid with minor semiaxis 150 kpc and major semiaxis 290 kpc for the shock volume, we obtain a shock energy of $E_{\text{sh}} = 6.5\pm2.0\times10^{61}$~erg. Over the range of shock ages $t_{sh} = 90 - 150$~Myr determined above, we find an average jet power of $(1.4 - 2.3)\times10^{46}$~erg~s$^{-1}$. These numbers are comparable with those presented in \citet{2005ApJ...625L...9N}, and place Hercules~A among the systems
with the highest jet power derived from shock fronts (e.g., \citealt{ubertosi23}). Moreover, the jet power vastly exceeds not only the bolometric X-ray luminosity in the innermost 100~kpc (where the gas cooling time is lower than 7.7~Gyr, see \citealt{2004MNRAS.350..865G,2005ApJ...625L...9N,2009ApJS..182...12C}) of $\sim$1.3~$\times~10^{44}$~erg~s$^{-1}$, but also the global cluster X-ray luminosity, $L_X\sim5\times10^{44}$~erg/s \citep{2004MNRAS.350..865G}, by nearly two orders of magnitude. Thus, the large outburst of Hercules~A is powerful enough to balance the radiative losses of the ICM due to cooling, as well as to stimulate turbulence in the ICM and cool gas inflows via chaotic cold accretion onto the SMBH (Singha et al. in prep.) Indeed, \citet{hofmann2016} find a (3D) turbulent Mach number of 0.21 in Hercules A, which is consistent with the typical turbulence injected via AGN jets in the ICM \citep{wittor2020}, thus providing another channel for non-thermal pressure and heating.

\vspace{0.2cm}
\par If the cavities were created by an older outburst, they should have larger dynamical ages than the cocoon shock front. The north-eastern and south-western X-ray cavities are located at 65 kpc and 70 kpc from the BCG of Hercules~A, respectively. Assuming that they have risen at the sound speed of the surrounding ICM ($\sim$1100~km~s$^{-1}$), their age is around $55 - 65$~Myr. Even accounting for the factor of $\sim$2 systematic uncertainties associated with dynamical ages of X-ray cavities (e.g., \citealt{2004ApJ...607..800B}), we find the timescale of the cavities to be consistent, at best, with the large scale cocoon shock front. 
\par Therefore, the origin of the X-ray cavities is unclear. The comparison with Cygnus A offers again a possible solution: the inner region of the cocoon shock around Cygnus A contains older radio plasma coincident with X-ray cavities. This older plasma might represent a backflow of particles from the lobes of the active jets, explaining its fainter emission and steeper spectral index \citep{2012A&A...545L...3C,2014ApJ...783...42M,2018MNRAS.476.4848D}. Hercules~A resembles Cygnus~A in terms of outburst power and cocoon shock morphology; if the X-ray cavities in Hercules~A are analogous to the backflows of Cygnus A, the electron population is old and the synchrotron emission would only be seen at low frequencies. Critically, if the cavities originated from backflows, the previously estimated age of 55 -- 65~Myr would be invalid. This is because the cavities would not have traveled from the cluster center to their current location at the sound speed but formed locally. Following this argument, we can compute the time that the cavities have taken by expanding to their present radius at the sound speed of the surrounding ICM. For cavity radii of $R\sim 45 - 65$~kpc (see Section \ref{subsec:xraycavities}), we obtain expansion timescales of $t_{\text{exp}} = R/c_{\text{s}} \sim 40 - 60 $~Myr. 
{For consistency, we can derive an approximate estimate of the timescale over which the backflowing bubble developed}, supposing that the plasma started to reflect backwards from the outer boundary of the lobes $t_\text{bf}$ ago, and afterwards the plasma expanded both forwards at roughly the speed of the leading shocks, $v_\text{sh}$ ($\sim$2000~km/s), and backwards at a sub-sonic speed $v_\text{bf}$. The extent of radio emission along the radio axis, $d_\text{pl}$ ($\sim$250~kpc), should statisfy:
\begin{align}
& d_\text{pl} = v_\text{sh}\cdot t_\text{bf}+v_\text{bf}\cdot t_\text{bf}\,, \quad\text{that is:}
\\
 &    t_\text{bf} = \frac{d_\text{pl}}{v_\text{sh} + v_\text{bf}}\,\,.
\end{align}
{We approximate the speed of expansion of the backflowing bubble with the ICM sound speed around the cavities ($v_\text{bf}\sim1100$~km s$^{-1}$)}, which gives $t_\text{bf}~\sim~75$~Myr. The fact that $t_\text{bf}\gtrsim t_\text{exp}$ confirms that the cavities had enough time to expand to their current sizes.
\par The partial overlap or absence of radio emission at 1.4 GHz covering the X-ray cavities (see Fig. \ref{fig:cuts}) aligns with the possibility of faded synchrotron radiation due to the electron population's energy losses. Any residual radio emission would likely be most evident at low frequencies. Indeed, there is barely detected, amorphous radio emission in the region of the X-ray cavities at 74~MHz and 144~MHz \citep{2005MNRAS.358.1061G,2022A&A...658A...5T}. As noted by \citet{2022A&A...658A...5T}, such diffuse radio emission has a spectral index steeper than $\alpha_{\text{diffuse}} = 1.8$. Further insights may come from high spatial resolution and sensitive radio observations at MHz frequencies (e.g., using LOFAR at 42-66 MHz with international stations). As a possible alternative, we briefly mention AGN-driven outflows or winds as a potential mechanism to explain radio-faint X-ray cavities. Simulations show that outflows can mechanically excavate bubbles in the ICM of galaxy clusters (e.g. \citealt{2011MNRAS.411..349G,prunier2024}). Observational estimates suggest mechanical powers of $10^{40}-10^{45}$~erg/s for winds, comparable to the typical power of jet-driven shocks and cavities \citep{2013MNRAS.430.1102T}. While initially associated with radio-quiet AGN, more recent evidence suggests that they are present also in radio-loud systems {\citep{2014MNRAS.443.2154T,2024MNRAS.532.3036M}}. If these winds can excavate cavities, the resulting bubbles would not be associated with synchrotron emission. The low excitation emission lines in the core of Hercules~A suggest that the SMBH is currently accreting at a low rate and would unlikely power a radiation-driven wind \citep{capetti2011}. If a wind was present, it must have been driven during a previous epoch when the SMBH was accreting at a higher level and with an ejection axis misaligned by $\sim$60$^{\circ}$ from the current one. 

\subsection{Non thermal radio emission along the jets and lobes of Hercules~A}\label{subsec:disc-nonthermal}
{
Besides the evident diffuse X-ray emission from the hot ICM, in the previous sections we found evidence for non-thermal emission from both the eastern jet and the radio lobes (Section \ref{subsubsec:xrayknot} and \ref{subsec:lobesIC}, respectively). Here we discuss the implication of those detections on the acceleration mechanisms, magnetic fields and particle content of the jets and lobes.
\subsubsection{The eastern X-ray jet: synchrotron or Inverse Compton emission?}\label{subsubsec:eastjetmodels}
Based on the best-fit model for the eastern X-ray jet discussed in Section \ref{subsubsec:xrayknot}, we find that the best-fit photon index of the power-law corresponds to a spectral index of $\alpha_{X} = \Gamma -1 = 0.93^{+0.34}_{-0.31}$,  
and we measure a 1~keV flux density of about 4.8~nJy from the X-ray jet. To understand which mechanism is responsible for the non-thermal X-ray emission, we consider the radio flux density of the jet within the green box shown in Fig. \ref{fig:easternjet}, measured from the 1.5, 4.9, 8.5 and 14.9 GHz VLA data at matching resolution of $\sim$1.5$''$. The flux densities at these frequencies are equal to $6.0\pm0.6$~Jy, $2.4\pm0.2$~Jy, $1.48\pm0.15$~Jy, and $0.88\pm0.09$~Jy respectively. The radio spectrum steepens across this frequency range, with $\alpha^{4.8}_{1.4} \sim 0.7 $ and $\alpha^{14.9}_{8.5} \sim 0.9$. 
To model the electron energy spectrum from the jets of Hercules~A we use the code \texttt{SYNCH}\footnote{\url{https://github.com/mhardcastle/pysynch}.} from \citet{1998MNRAS.294..615H}. We describe the volume of the jet using a cylinder with length of 20$''$ and radius 3$''$, and we use a broken power-law to model the electron energy spectrum. We further assume a an electron energy index $p = 2.2$ and a minimum Lorentz factor of $\gamma_{min} = 10$. The choice of $p =2\alpha_{i}+1= 2.2$ is justified by the injection index $\alpha_{i} = 0.62$ measured for Hercules~A \citep{2003MNRAS.342..399G,2005ApJ...626..748Y}, which is close to the assumption of first-order Fermi particle acceleration, $p=2$ (e.g., \citealt{2010MNRAS.404.2018H}). As visible in Fig. \ref{fig:easternjet}c, the radio spectrum is well described by the broken power-law with a Lorentz factor of the break $\gamma_{b} = 9.5\times10^{3}$ and slope $\alpha_{\gamma>\gamma_{b}} = 0.93$. The ``equipartition'' or ``minimum energy'' methods of \texttt{SYNCH} return consistent values of the magnetic field, between $36-40$~$\mu$G. 
\par Two possible explanations for the X-ray emission from the jet are either synchrotron emission up to the X-ray band, or IC. The first option requires high maximum Lorentz factors $\gamma_{max}$ of the electron population, up to a few $10^{7}$. Local re-acceleration would also be necessary for the X-ray synchrotron emission to be visible, because the lifetime of electrons with $\gamma_{max}\sim10^{7}$ is $10^{3}-10^{4}$~yr for $\mu$G-level magnetic fields (e.g., \citealt{2009A&ARv..17....1W}). The blue and cyan lines in Fig. \ref{fig:easternjet}c show the extrapolation of the model for the radio spectrum at high frequencies for $\gamma_{max} = 10^{5}$ and $\gamma_{max} = 2\times10^{7}$, respectively. Clearly, high Lorentz factors under-predict the X-ray fluxes by orders of magnitude with respect to our measurement of 4.8~nJy. Even higher Lorentz factors (above $\gamma_{max}\geq 10^{8}$) would theoretically be possible, but it would necessitate extremely efficient re-acceleration of electrons at $\sim$100 kpc from the center. This re-acceleration may be provided by e.g., internal shocks at the location along the jets where the X-ray emission from the jet appears. For $\mu$G-level magnetic fields along the jet and internal shock speed $\sim$0.1$c$, the characteristic timescale predicted by diffusive shock acceleration to accelerate the electrons to $\gamma_{max}\sim10^{8}$ is $10-100$~yr (e.g., equation 1 in \citealt{2003ApJ...585L.113B}), which is less than the electrons' synchrotron lifetime. Future investigation of the eastern jet emission in the optical wavelength, where the synchrotron scenario predicts a flux density of $\sim$0.01 -- 0.1~mJy, may reveal if in-situ acceleration of electrons is at the origin of the X-ray emission. 
\par The second option is the upscattering of CMB photons to X-ray energies by electrons with $\gamma\sim10^{3}-10^{4}$. A possible evidence in favour of IC comes from the slope of the X-ray power-law: if the X-rays had a synchrotron origin, one would expect to find a steeper spectrum at X-ray frequencies, whereas the IC mechanism produces a spectrum identical to the synchrotron one but shifted to higher energies. The $\alpha_{X} = 0.93^{+0.34}_{-0.31}$ of the jet in Hercules~A is consistent with the radio slope, even though the large errorbars mean that this constraint is not very robust. Fig. \ref{fig:easternjet}c shows the predictions of the IC-CMB flux density based on the radio spectrum model (gray dotted line). The predicted flux density at 1~keV falls short by a factor of $\sim$30 with respect to the observed one. This apparent mismatch can be resolved if considering that the IC emission can be Doppler-boosted, because the electrons ``see'' a beamed CMB flux boosted by a factor $\delta^{3+p}$, where $\delta$ is the Doppler factor and $p$ is the electron energy index \citep{2001ApJ...561..111G}. By considering the dependencies of the synchrotron and IC flux densities on the Doppler factor (see e.g., \citealt{1995PASP..107..803U,2001ApJ...561..111G,2010A&A...516A..18D}) we find that the predicted and observed IC flux densities can be reconciled with a moderate Doppler boosting factor of $\delta\sim2.7$ and a lower equipartition magnetic field $B\sim12\mu$G (black dashed line in Fig.~\ref{fig:easternjet}c). The brightness variations across the twisted radio shape of the eastern X-ray jet are consistent with appreciable Doppler boosting occurring at different sites across the spiraling jet path (see also \citealt{2003MNRAS.342..399G}). 
\par To conclude, we note that the turn-on of the X-ray emission from the jet spatially follows the location along the jet at which decollimation takes place (knot E7 in \citealt{2003MNRAS.342..399G}). The radio images reveal that the jet bends and widens at about 73~kpc in projection from the center \citep{2003MNRAS.342..399G,2022A&A...658A...5T,perucho2023}, and then produces the twisting structures visible in Fig. \ref{fig:easternjet}. Thus, in the east direction, the jet not only undergoes decollimation but also deflection. Identifying any potential obstacles along the jet's path is beyond the scope of this paper. However, it is worth noting that the intrinsic absorbing column density over the surface area of the eastern jet region (see Section \ref{subsubsec:xrayknot}) implies a total absorbing mass of $3 - 14 \times10^{8}$~M$_{\odot}$, for a volume filling factor of 0.1 - 0.5 (e.g., \citealt{1997MNRAS.286..583A}). The absorbing gas may have been stripped from the object that deflected the jet; alternatively, it may have been uplifted from the central galaxy to tens of kpc by the jets due to entrainment (see \citealt{2013ApJ...771...38O}). Given the highly speculative nature of this hypothesis, we will not delve into it further.

\subsubsection{Magnetic field and particle content of the large radio lobes in Hercules~A}\label{subsubsec:magfieldlobes}
The IC flux density from the radio lobes can be combined with the radio synchrotron emission to constrain the magnetic field within the lobes (see e.g., \citealt{1998MNRAS.294..615H,2000ApJ...544L..23T,2009A&ARv..17....1W,2016ApJ...826..109C}). Synchrotron emission depends on both the electron distribution and the magnetic field, while the IC emission depends on the electron distribution only (e.g., for a review \citealt{1970RvMP...42..237B}). Therefore, using measurements of radio synchrotron and corresponding X-ray IC emission, it is possible to estimate the average magnetic field. Following \citet{2019MNRAS.486.5430M}, we consider that the total X-ray emission from the radio lobes of Hercules~A at {1~keV} is $21.7\pm 1.4 \,\text{(statistical)}\pm 1.3 \,\text{(systematic)}$~nJy. The total synchrotron flux density at 1.4 GHz from the same region is $13.9\pm0.9$~Jy. The total radio spectrum of Hercules A, dominated by the emission from the radio lobes, is steep with $\alpha \sim1.2$ down to 200 MHz \citep{2017ApJS..230....7P}, and the $12.6-25$~MHz spectral index $\alpha\approx1.0$ \citep{1969MNRAS.143..289B} implies that the spectrum remains steep down to tens of MHz. We thus adopt a spectral index $\alpha = 1.2$ for the lobes of Hercules~A. Using equation 6 in \citet{2019MNRAS.486.5430M}, we estimate a volume-average magnetic field of $B \sim 12\pm3$~$\mu$G. Values of a few $\mu$G are typical of the lobes of radio galaxies with detection of or upper limits to extended IC emission (e.g., \citealt{feigelson1995,brunetti1999,hardcastle2002}). The above $B$ may be interpreted as a lower limit on the true magnetic field, because if a fraction of the X-ray emission is not IC of the CMB photons, the particle content must be lower, requiring a stronger magnetic field to produce the same synchrotron flux density. Any such fraction is likely small, as we excluded a major contribution from thermal plasma (Appendix \ref{app:thermlobes}), and the IC scattering of synchrotron photons is negligible, as we demonstrate here. The photon energy density of the CMB is $u_\text{CMB} = 4.2\times10^{-13}\times(1+z)^{4}\sim 7.5\times10^{-13}$~erg~cm$^{-3}$. The synchrotron photon energy density is $u_\text{syn} = 3L_\text{syn}/4\pi cR^{2}$ (e.g., \citealt{2010ApJ...714...37Y}), where $L_\text{syn}$ is the integrated synchrotron luminosity and R is the radius of the lobe ($\sim$90~kpc). $L_\text{syn}$ was calculated by integrating the spectrum of the lobes between 10 MHz and 100 GHz, assuming a power-law spectrum, $S_\nu = S_{\nu_0}(\nu/\nu_0)^{-\alpha}$, with $S_{\nu_0} = 13.9$~Jy at $\nu_0 = 1.4$~GHz and $\alpha = 1.2$. We find $u_\text{syn} = 1.3\times10^{-14}$~erg~cm$^{-3}$, which is clearly negligible compared to the CMB.
\par Using the constraint on the lobe pressure given by the shocked ICM around the lobes, we can study the particle content of the radio lobes in Hercules~A (see for comparison e.g., \citealt{2005ApJ...626..733C,2010MNRAS.404.2018H,2016MNRAS.458.4443H}). We consider that the pressure of the lobes, $p_{lobe}$ should be equal to the average pressure of the shocked ICM around the radio lobes, of $p_{\text{ICM}} = 5.4\pm0.3\times10^{-11}$~erg~cm$^{-3}$ (see Table \ref{tab:specfronts}). The pressure inside the lobes is given by the sum of the contributions from the magnetic field, the electrons, and the nonradiating particles, so that:
\begin{equation}
    p_{\text{ICM}} \approx p_{lobe} = p_{B} + p_{e} + p_{n}
    \label{eq:pressurecontrib}
\end{equation}
\noindent {The magnetic pressure can be derived from the magnetic field as $p_{B} = B^{2}/24\pi = 1.9\pm0.8\times10^{-12}$~erg~cm$^{-3}$.} For a given assumption about the electron energy spectrum, we can estimate the internal pressure of the lobe provided by electrons. We model the electron energy spectrum from the lobes of Hercules~A assuming 
{a broken power-law with $\alpha_{\gamma<\gamma_{b}} = 0.62$, $\gamma_{b} = 2.5\times10^{2}$ ($\sim$10~MHz), $\alpha_{\gamma>\gamma_{b}} = 1.2$, minimum and maximum Lorentz factors $\gamma_{min} = 10$ and $\gamma_{max} = 10^{5}$, and considering the magnetic field of $B=12\pm3$~$\mu$G determined above. 
We find that the electron pressure is $p_{e} = 3.6\pm1.2\times10^{-11}$~erg~cm$^{-3}$. The comparison between $p_{B}$ and $p_{e}$ confirms the prediction from \citet{2010MNRAS.404.2018H} that the lobes are not magnetically dominated. The ratio between $p_{\text{ICM}}$ and $p_{B}+p_{e}$ of $1.4\pm0.5$ is consistent, within the large uncertainty, with nonradiating particles contributing less than electrons to the total lobe pressure.}

\section{Summary}\label{sec:conclusion}
We summarize here our findings on Hercules~A:
\\ \textbullet\,\, 
We detected east and west shock fronts at 280~kpc from the center of Hercules~A, forming part of the \ul{cocoon shock front} surrounding the central radio galaxy. The cocoon shock has a higher Mach number to the east and west, $\mathcal{M}_{\text{SB}} = 1.9\pm0.3$, than to the north and south, $\mathcal{M}_{\text{SB}} = 1.65\pm0.05$, compatible with the orientation of the jet propagation axis. 
Based on dynamical arguments, we estimate that the age of the cocoon shock lies between 90 and 150 Myr, which gives an indication of the age of the large radio lobes.
\\ \textbullet\,\, 
We confirmed the presence of two \ul{radio-faint X-ray cavities} of $\sim$50~kpc in radius and at $\sim$70~kpc from the center.
A backflow from the large radio lobes may explain why the X-ray cavities are dynamically younger (40 - 60 Myr) but any associated radio emission is radiatively older than the active radio lobes. Alternatively, the cavities may have been excavated by AGN-driven winds, explaining their unclear connection with synchrotron emission.
\\ \textbullet\,\, 
The \ul{X-ray emission along the eastern jet} of Hercules A, located at approximately 80 kpc (in projection) from the center, is best fit by an intrinsically absorbed power-law with a photon index $\Gamma \sim 1.9$ and flux density of $\sim$4.8~nJy at 1~keV.
We considered two scenarios for the jet's X-ray emission: high-energy synchrotron, which would need highly efficient re-acceleration of electrons with $\gamma \sim 10^8$ at $\sim$80 kpc from the center, and IC scattering of the CMB, requiring Doppler boosting $\delta \sim 2.7$ and a magnetic field $B \sim 12~\mu$G.
\\ \textbullet\,\, 
We detected \ul{X-ray IC emission from the radio lobes of Hercules~A} from surface brightness and spectral analysis. This non-thermal emission explains why there are no visible cavities associated with the powerful lobes of Hercules~A. The 1 keV flux density of the two lobes together is $21.7\pm1.4~\,\text{(statistical)}\pm~1.3\,\text{(systematic)}$~nJy. Combined with the co-spatial synchrotron flux density, the IC detection yields a volume-average magnetic field of $12\pm3$~$\mu$G.} 

\vspace{0.2cm}
\par Overall, our analysis revealed new insights into the interaction between the peculiar central radio galaxy of Hercules~A and the surrounding hot gas. A companion work will also address the relation between the AGN outburst and the central warm gas nebula cooling out of the X-ray halo in the central tens of kpc (Singha et al. in prep.). However, several questions on the large-scale environment remain that require deeper X-ray observations for definitive answers. One key unknown is the exact dynamic history of the huge cocoon shock front, knowledge of which is currently limited by the relatively low X-ray statistics at the large distance (280 kpc) of the east-west edges from the center. Very deep {\it Chandra} data would be crucial to dissect the properties, dynamics, and origin of  Hercules~A cocoon shock, which is itself intriguing due to its 4-fold larger size compared to the best-studied Cygnus A (for which $\sim$2 Ms of {\it Chandra} data exist). High-quality spatial and spectral X-ray data would be necessary to further investigate the inverse Compton (IC) emission from the jet and radio lobes. This could potentially involve spatially resolved mapping of the non-thermal emission. Ultimately, hard X-ray observations would be vital to isolate the non-thermal X-ray emission from the lobes and definitely distinguish IC emission from the hard X-ray spectrum.

\begin{acknowledgements}
This research has made use of {\it Chandra} datasets, obtained by the {\it Chandra} X-ray Observatory, contained in the {\it Chandra} Data Collection (CDC): \url{https://doi.org/10.25574/cdc.286}. The radio (NVAS) images shown in this paper were produced as part of the NRAO VLA Archive Survey, (c) AUI/NRAO. S.B. and C.O. acknowledge support from the Natural Sciences and Engineering Research Council (NSERC) of Canada. M.G. acknowledges support from the ERC Consolidator Grant \textit{BlackHoleWeather} (101086804).
\end{acknowledgements}

\bibliographystyle{aa}
\bibliography{aanda}

\begin{appendix}
\onecolumn
\section{Details on the spectral analysis of the shock fronts}\label{app:details}
\begin{figure*}[ht!]
   \centering
   \sidecaption
   \includegraphics[width=0.7\linewidth]{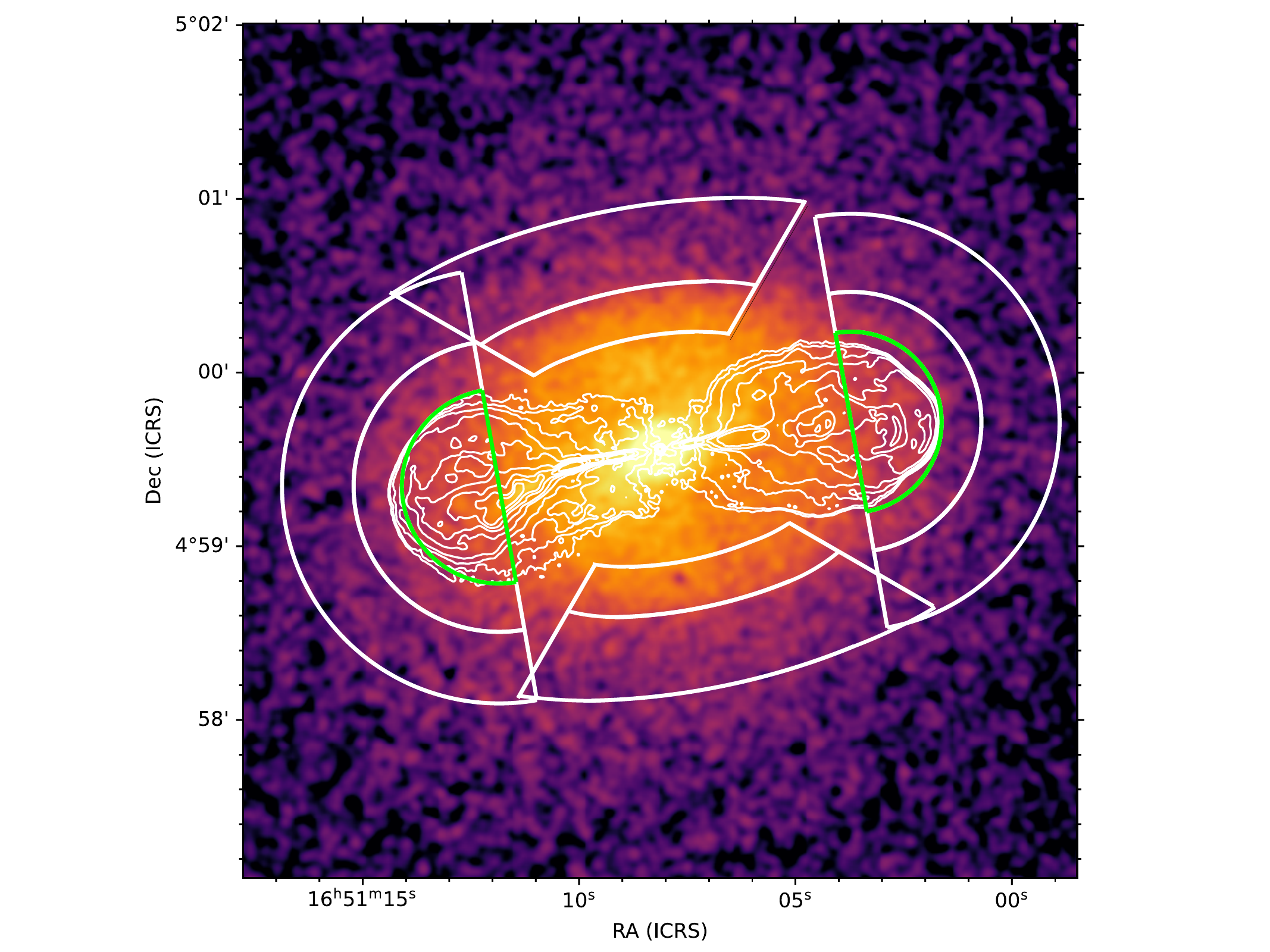}
      \caption{Downstream and upstream regions used for the spectral analysis of the north-south and east-west shock fronts in Hercules~A (white wedges), overlaid onto the 0.5 -- 7 keV {\it Chandra} image of the cluster. A third outer wedge extending to the edge of the field of view was used to deproject the spectra (not shown here). The green semicircles mark the regions used for the spectral analysis of IC emission from the radio lobes. Radio contours are the same as in Fig.~\ref{fig:shockboundary}. See Section \ref{subsec:specshocks} and \ref{subsec:lobesIC} for details.}\label{fig:specanalysisregions}
   \end{figure*}

We report in Table \ref{tab:specfronts} the details of the spectral fits to the spectra of the wedges shown in Fig. \ref{fig:specanalysisregions}, that we used to study the spectral properties of the surface brightness edges detected in Section \ref{subsec:specshocks}. A third outer wedge extending to the edge of the field of view was used to deproject the spectra (not shown here).

\setlength{\tabcolsep}{3pt}
\begin{table*}[ht!]
		\centering
		\caption{Spectral analysis of the surface brightness edges in Hercules~A. 
        }
		\renewcommand{\arraystretch}{1.5}
		\label{tab:specfronts}
		\begin{tabular}{c|c|c|c|c|c|c|c|c}
\hline
			 & C-stat/D.o.f. & Side & Net counts & R$_{\text{i}}$ & R$_{\text{o}}$  & $kT$ &    $p_{\text{ICM}}$ & $n_{e}$\\
			& & & & [kpc ($''$)] & [kpc ($''$)] & [keV] &   [10$^{-11}$ & [10$^{-3}$ \\
            & & & &  &  &  &   erg cm$^{-3}$] & cm$^{-3}$]\\

\hline
  		\multirow{2}{*}{North-South$^{a}$} & \multirow{2}{*}{5507.8/5320 (1.04)} & \textit{downstream} & 8296 & 104.6 (39.4) & 149.5 (56.3) & 5.68$^{+0.29}_{-0.33}$ (5.01$^{+0.26}_{-0.24}$)  &  6.89$^{+0.41}_{-0.46}$ & 4.13$^{+0.12}_{-0.12}$\\

		 & & \textit{upstream} & 4680 & 149.5 (56.3) & 224.0 (84.3)  & 4.01$^{+0.19}_{-0.21}$ (3.63$^{+0.16}_{-0.18}$) & 2.42$^{+0.12}_{-0.13}$ & 2.06$^{+0.04}_{-0.04}$\\

\hline

		\multirow{2}{*}{East}  & \multirow{2}{*}{627.2/573 (1.09)} & \textit{downstream} & 1131 & 94.5 (35.4) & 139.4 (52.2) & 7.74$^{+1.63}_{-1.15}$ (7.10$^{+0.92}_{-0.82}$)  &  5.67$^{+1.23}_{-0.90}$ & 2.49$^{+0.14}_{-0.14}$\\

		 & & \textit{upstream} & 1780 & 139.4 (52.2) & 208.3 (78.0)  & 5.72$^{+1.33}_{-0.94}$ (6.15$^{+1.13}_{-0.79}$) & 2.40$^{+0.47}_{-0.43}$ & 1.43$^{+0.04}_{-0.04}$\\
		
\hline

		\multirow{2}{*}{West} & \multirow{2}{*}{511.9/475 (1.08)} & \textit{downstream} & 955 & 83.3 (31.2) & 120.2 (45.0) & 6.84$^{+2.70}_{-1.73}$ (5.58$^{+1.12}_{-0.84}$)  &  5.05$^{+2.00}_{-1.29}$ & 2.51$^{+0.08}_{-0.08}$\\

		 & & \textit{upstream} & 1391 & 120.2 (45.0) & 192.2 (72.0)  & 4.14$^{+1.06}_{-0.87}$ (4.08$^{+0.59}_{-0.51}$) & 1.46$^{+0.40}_{-0.33}$ & 1.20$^{+0.10}_{-0.09}$\\
  
\hline
		\end{tabular}
		\tablefoot{(1) Name of the edge; (2) C-stat/D.o.f.; (3) side of the edge; (4) Number of net counts; (5-6) inner and outer radius of the wedge. Note that the center of the inner wedges is the center of Hercules~A, while the centers of the east and west outer wedges is the center of the associated radio lobe; (7) deprojected temperature (projected temperature inside brackets); (8) deprojected pressure; (8) deprojected electron density.
  \\$^{a}$ The inner and outer radii are relative to the minor axis of the elliptical wedges with ellipticity 1.75 and position angle 10$^{\circ}$ shown in Fig. \ref{fig:specanalysisregions}. }
\end{table*}

\section{Testing a thermal model for the X-ray emission within the radio lobes}\label{app:thermlobes}
In Section \ref{subsec:lobesIC} we reported on the spectral analysis of the X-ray emission from the lobes of Hercules~A, that we interpret as non-thermal emission caused by IC. In this Appendix, we explore the alternative hypothesis that the X-ray emission from the lobes of Hercules~A is of thermal origin. We used the same spectral extraction region of Section \ref{subsec:lobesIC}. In order to properly fit the spectra, we considered again two possibilities for the background:
\begin{enumerate}[leftmargin=*]
    \item We adopted as background the shell of shocked gas immediately outside the radio lobe, whose spectrum accounts for the whole emission projected onto the lobe. We fitted the spectra with an absorbed \texttt{apec} model in the form \texttt{tbabs$\ast$apec}, leaving the normalization and temperature of the \texttt{apec} component free to vary and fixing the abundance to $Z=0.5$~Z$_{\odot}$ (as measured in an off-lobe region, see Section \ref{subsec:lobesIC}). We find, for the spectrum of the eastern lobe, a temperature $kT = 8.3\pm2.1$~keV, and a normalization of $(9.6\pm0.8)\times10^{-5}$ (C-stat/D.o.f. = 259.9/266). For the spectrum of the western lobe we find a temperature $kT = 7.3\pm1.6$~keV, and a normalization of $(6.4\pm0.7)\times10^{-5}$ (C-stat/D.o.f. = 196.8/217). 
    \item As an alternative approach, we subtracted the blank-sky background from the source spectrum. Since the blank-sky does not account for thermal emission within or projected onto the lobes, 
    we fitted the lobe spectrum with a \texttt{tbabs$\ast$(apec$+$apec)} model. As done in Section \ref{subsec:lobesIC}, the temperature and abundance of the first thermal model, which represents the emission from the gas projected onto the lobe, were frozen to $kT = 4.1$~keV and $Z=0.5$~Z$_{\odot}$ (see Section \ref{subsec:lobesIC}). The abundance of the second thermal model was also frozen to $Z=0.5$~Z$_{\odot}$. For the eastern lobe spectrum we find a temperature $kT = 8.4\pm3.1$~keV, and a normalization of $(7.0\pm0.9)\times10^{-5}$ (C-stat/D.o.f. = 262.1/226).
    For the western lobe we find temperature $kT = 8.6\pm3.6$~keV, and a normalization of $(3.5\pm0.8)\times10^{-5}$ (C-stat/D.o.f. = 171.7/184). 
\end{enumerate}
Similarly to the non-thermal case of Section \ref{subsec:lobesIC}, we find consistent results from the two methods. We note that the best-fit temperatures for the thermal plasma supposedly filling the radio lobes are higher than those of the shocked gas immediately outside the lobes (even though the uncertainties are relatively large). This indicates that the spectra are fairly hard, supporting our favored conclusion that the spectra are best described by a power-law component representing non-thermal emission (Section \ref{subsec:lobesIC}).
\\{Additionally, if the lobes were indeed filled with thermal plasma, their pressure, $p_{lobe}^{therm}$, should be comparable to or lower than the pressure of the downstream shocked gas, $p_{\text{ICM}}^{down}$ (since non-thermal pressure from synchrotron emission would also contribute). We test this scenario by determining the electron density inside the lobes, in the hypothesis that they are filled with ICM. While we did not perform a classic deprojection analysis to derive the normalizations above, our choices of background are meant to subtract the contribution from the gas projected onto the line of sight. Thus, for case (1) reported above, we obtain electron densities of $n_{e} = (3.4\pm0.1)\times10^{-3}$~cm$^{-3}$ and $n_{e} = (3.3\pm0.1)\times10^{-3}$~cm$^{-3}$ for the eastern and western lobes, respectively. For case (2) we obtain electron densities of $n_{e} = (2.9\pm0.1)\times10^{-3}$~cm$^{-3}$ and $n_{e} = (2.5\pm0.1)\times10^{-3}$~cm$^{-3}$ for the eastern and western lobes, respectively. For comparison, the emission per unit volume $A_l$ of each lobe determined in Section \ref{subsubsec:sbprofiles} can be converted to an electron density assuming a temperature and abundance for a thermal plasma. Using an \texttt{apec} model with $5\leq kT\leq8$~keV and $Z=0.5$~Z$_{\odot}$ (see Section \ref{subsec:lobesIC}), the $A_l$ of the eastern and western lobes correspond to $n_e\sim3.5\times10^{-3}$~cm$^{-3}$ and $n_e\sim3.8\times10^{-3}$~cm$^{-3}$, respectively. All of these estimates for the ICM density are higher than the density of the plasma in the shocked gas immediately outside the radio lobes (see Table \ref{tab:specfronts}). From the derived electron density and temperature, we estimate the pressure of the thermal plasma supposedly filling the lobes to be $\sim 7 - 10 \times 10^{-11}$~erg~cm$^{-3}$. This yields $p_{lobe}^{therm}\sim 1.5 - 2 \times p_{\text{ICM}}^{down}$. This supports the interpretation that the lobes are dominated by non-thermal plasma, as the thermal plasma hypothesis leads to pressures that are inconsistent with the surrounding environment.}
\\Furthermore, the thermal model would have non-physical implications on the polarization properties of the radio lobes. The synchrotron emission from the radio lobes of Hercules~A is polarized at GHz frequencies, with polarization fractions $S_{\nu,pol}^{obs}/S_{\nu,tot} \sim 30\% - 40\%$ at $\nu = 4.8$~GHz and $\nu = 8.4$~GHz (figures 12 - 16 in \citealt{2003MNRAS.342..399G}). If the density of thermal plasma inside the radio lobes was very high, the lobes should be completely depolarized. Faraday rotation of synchrotron emission passing through a {uniformly} magnetized plasma can be described in terms of rotation measure, $\text{RM}$:
\begin{equation}
    \text{RM[rad/m$^{2}$]} = 812\int_{0}^{L} n_{e}\text{[cm$^{-3}$]}\,B_{\parallel}\text{[$\mu$G]} \,dl\text{[kpc]}
\end{equation}
where $L$ is the path length through the plasma and $B_{\parallel}$ is the component of the magnetic field along the line of sight \citep{2004IJMPD..13.1549G}. Due to Faraday depolarization, the observed polarized flux density at wavelength $\lambda$ is lower than the intrinsic one by a factor:
\begin{equation}
    \frac{S_{\nu,pol}^{obs}}{S_{\nu,pol}^{int}} = \frac{\sin(2\text{RM}\lambda^{2})}{2\text{RM}\lambda^{2}}.
\end{equation}
The intrinsic fraction of polarization of synchrotron emission depends on the injection index $p$ of the electron population, and can reach $S_{\nu,pol}^{int}/S_{\nu,tot} \sim 70\% - 75\%$ for typical $p = 2,2.5,3$  (e.g., \citealt{1980MNRAS.193..439L}). Based on the above equations, for electron densities between $2.5 - 3.8\times10^{-3}$~cm$^{-3}$ and a typical magnetic field of 5~$\mu$G, we find a rotation measure $\text{RM} \sim 1000 - 1400$~rad/m$^{2}$. We thus estimate the expected depolarization fractions, finding $S_{\nu,pol}^{obs}\leq 10\%S_{\nu,pol}^{int}$ at 4.8~GHz and $S_{\nu,pol}^{obs}\leq 30\% S_{\nu,pol}^{int}$ at 8.4~GHz. Given the observed polarization fractions $S_{\nu,pol}^{obs}/S_{\nu,tot} \sim 30\% - 40\%$, we would obtain an intrinsic polarized flux density exceeding the maximum theoretical value of $75\%S_{\nu,tot}$, reaching values greater than $100\%S_{\nu,tot}$, which is clearly nonphysical. If the lobes were indeed filled with a thermal plasma of electron density $n_e\sim3\times10^{-3}$~cm$^{-3}$, the intrinsic degree of polarization would remain below the threshold of $75\%S_{\nu,tot}$ only if the magnetic field was very low, about $\leq1$~$\mu$G. For completeness, we note that these calculations assume a line-of-sight uniform magnetic field within the lobe. If, instead, the magnetic field was primarily randomized due to turbulence, the internal rotation measure would be reduced by a factor $\sqrt{L/d}$, where $L$ is the depth through the lobe and $d\ll L$ is the coherence length. Consequently, depolarization would be significantly reduced \citep{1966MNRAS.133...67B,1998MNRAS.299..189S}. 
\\Given the above considerations on the temperature, density, and pressure of any thermal component inside the lobes, and the calculations on the expected polarization fractions, we favour the non-thermal IC interpretation of the X-ray emission (Section \ref{subsec:lobesIC} and \ref{subsubsec:magfieldlobes}).
\end{appendix}

\end{document}